\documentclass[12pt]{article}
\usepackage[english]{babel}
\usepackage{url}
\usepackage[utf8x]{inputenc}
\usepackage{amsmath}
\usepackage{graphicx}
\graphicspath{{images/}}
\usepackage{parskip}
\usepackage{fancyhdr}
\usepackage{vmargin}
\usepackage{adjustbox}
\usepackage{color}
\usepackage{graphicx,rotating,booktabs}

\usepackage[verbose]{placeins}

\setmarginsrb{1.5 cm}{0 cm}{1.5 cm}{0.5 cm}{1 cm}{1.5 cm}{1 cm}{1.5 cm}

\title{Essential Motor Cortex Signal Processing: an ERP and functional connectivity MATLAB toolbox - user guide version 2.0}                             
\author{Esmaeil Seraj}                               
\date{\today}                                           

\makeatletter
\let\thetitle\@title
\let\theauthor\@author
\let\thedate\@date
\makeatother

\begin{document}


\begin{titlepage}
    \centering 
	\includegraphics[scale = 0.5]{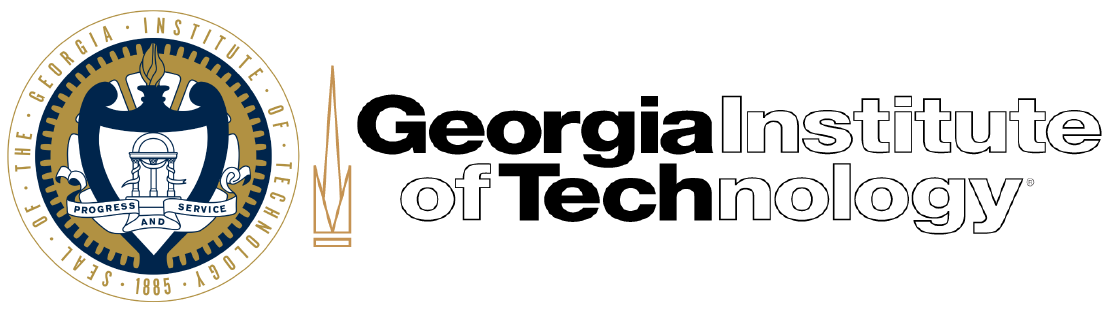}\\[1.0 cm]    
    \includegraphics[scale = 0.5]{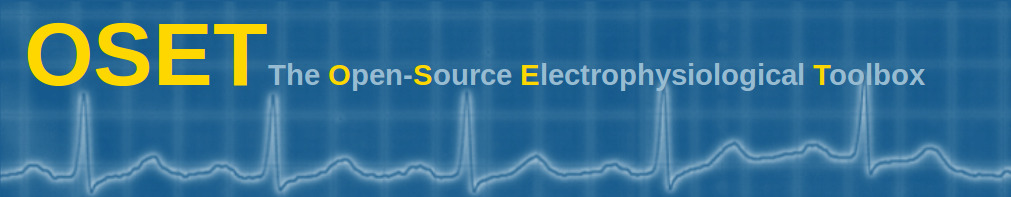}\\[3.0 cm] 
    \textsf{\Large Georgia Institute of Technology, Atlanta (GA), United States}\\[0.5 cm]               
    \textsf{\large School of Electrical and Computer Engineering}\\[1 cm]               
    \rule{\linewidth}{0.5 mm} \\[0.4 cm]
    { \Large \bfseries \thetitle}\\
    \rule{\linewidth}{0.5 mm} \\[1 cm]
    
    \begin{minipage}{0.5\textwidth}
        \begin{flushleft} \large
            Esmaeil Seraj$ ^{\dagger , *} $\\
            \end{flushleft}
            \end{minipage}~
            \begin{minipage}{0.5\textwidth}
            \begin{flushright} \large
            Karthiga Mahalingam$ ^{\dagger} $\\
             \hspace{2 cm}\\
        \end{flushright}
    \end{minipage}\\[0.5 cm]
    
    \rule{\linewidth}{0.1 mm}
    \begin{flushleft}
    {\footnotesize $ ^* $ Corresponding author}\\[0.1 cm]
    \end{flushleft}
    \begin{flushleft}
    {\footnotesize $ ^\dagger $Correspondences shall be forwarded to: \textit{electronic mail:} \{eseraj3, kmahalingam\}@gatech.edu}\\[0.5 cm]
    \end{flushleft}
    {\large \thedate}\\
 
    \vfill
    
\end{titlepage}


\vspace*{3.5 cm} 
\begin{abstract}
\noindent The purpose of this document is to help individuals use the "Essential Motor Cortex Signal Processing MATLAB Toolbox". The toolbox implements various methods for three major aspects of investigating human motor cortex from Neuroscience view point: (1) ERP estimation and quantification, (2) Cortical Functional Connectivity analysis and (3) EMG quantification. 

\noindent The toolbox – which is distributed under the terms of the GNU GENERAL PUBLIC LICENSE as a set of MATLAB\textregistered ~routines – can be downloaded directly at the address:

 \begin{center}
 \underline{\texttt{http://oset.ir/category.php?dir=Tools}}
  \end{center}.
 
 or from the public repository on GitHub, at address below:

 \begin{center}
 \underline{\url{https://github.com/EsiSeraj/ERP_Connectivity_EMG_Analysis}}
 \end{center}.

\noindent The purpose of this toolbox is threefold: 

\begin{enumerate}
\item Extract the event-related-potential (ERP) from preprocessed cerebral signals (i.e. EEG, MEG, etc.), identify and then quantify the event-related synchronization/desynchronization (ERS/ERD) events. Both time-course dynamics and time-frequency (TF) analyzes are included.

\item Measure, quantify and demonstrate the cortical functional connectivity (CFC) across scalp electrodes. These set of functions can also be applied to various types of cerebral signals (i.e. electric and magnetic).

\item Quantify electromyogram (EMG) recorded from active muscles during performing motor tasks.
\end{enumerate}
\vspace*{2cm}
\textbf{Key-words:} Event-Related Potential, ERP, Event-Related Synchronization, ERS, Event-Related Desynchronization, ERD, ERP Time Dynamics, Time-Frequency Analysis, Cortical Functional Connectivity, CFC, Electroencephalogram, EEG, Electromyogram, EMG, EMG Quantification, MATLAB Function, Free Toolbox, User Guide, Manual
\end{abstract}


\newpage
\tableofcontents
\pagebreak


\section{Introduction}
\label{sec:gettingstarted}
This document is meant to help individuals use the "Essential Motor Cortex Signal Processing MATLAB Toolbox". Using the most popular reference articles in literature, the toolbox implements various methods for three major aspects of neuro-physiological investigation of human motor cortex: (1) ERP estimation and quantification (e.g. based on \cite{pfurtscheller1999event, handy2005event, luck2014introduction, makeig2004mining}), (2) Cortical Functional Connectivity analysis (e.g. based on \cite{greicius2003functional, carter1973estimation, lachaux1999measuring, varela2001brainweb, rosenblum1996phase}) and (3) EMG quantification (e.g. based on \cite{ricamato2005quantification, walter1984temporal}). 

The purpose of this toolbox is threefold: 

\begin{enumerate}
\item Extract the event-related potential (ERP) from preprocessed cerebral signals (i.e. EEG, MEG, etc.), identify and then quantify the event-related synchronization/desynchronization (ERS/ ERD) events. Both time-course dynamics and time-frequency (TF) analysis are included.

\item Measure, quantify and demonstrate the cortical functional connectivity (CFC) across scalp electrodes. These set of functions can also be applied to various types of cerebral signals (i.e. electric and magnetic).

\item Quantify electromyogram (EMG) recorded from active muscles during performing motor tasks.
\end{enumerate}

A primary goal of this toolset is to ease understanding the routines and help individuals alter our codes according to their study and/or implement their own techniques. For this purpose, herein, we first present a detailed tutorial (e.g. Section~\ref{subsec:fundamentals}.) on signal preconditioning and interpretable-implementation of introduced methods, from a practical view-point.

\subsection{License - No Warranty}
\label{subsec:license}
This program is free software; you can redistribute it and/or modify it under the terms of the GNU GENERAL PUBLIC LICENSE as published by the Free Software Foundation; either version 2 of the License, or (at your option) any later version.

This program is distributed in the hope that it will be useful, but WITHOUT ANY WARRANTY; without even the implied warranty of MERCHANTABILITY or FITNESS FOR A PARTICULAR PURPOSE. See the GNU GENERAL PUBLIC LICENSE for more details. You should have received a copy of the GNU GENERAL PUBLIC LICENSE along with this program; if not, see $ \langle $ \underline{\texttt{http://www.gnu.org/licenses/}} $ \rangle $ or write to the Free Software Foundation, Inc., 51 Franklin Street, Fifth Floor, Boston, MA  02110-1301, USA.

\subsubsection{The Open-Source Electrophysiological Toolbox (OSET)}
\label{subsub:oset}
Open Source Electrophysiological Toolbox (OSET) is a collection of electrophysiological data and open source codes for biosignal generation, modeling, processing, and filtering. OSET, version 3.1, 2014 Released under the GNU GENERAL PUBLIC LICENSE \cite{sameniopen}. Copyright\textcopyright ~2012.

As a progressive general-purpose open-source toolset, OSET is one of the main sources to access the functions and codes of current \textit{motor cortex signal processing} toolbox and also its documentation and dependencies.

\subsection{Citation}
\label{subsec:citation}
Within the limits of the GNU GENERAL PUBLIC LICENSE, you can use the toolbox as you please; however, if you use the toolbox in a work of your own that you wish to publish, you need to make sure to cite this user manual and the original studies properly, as shown below. This way you will contribute to helping other scholars find these items.

\begin{itemize}

\item Esmaeil Seraj and Karthiga Mahalingam “Essential Motor Cortex Signal Processing: an ERP and functional connectivity MATLAB toolbox - User Guide Version 1.0,” arXiv Preprint, June 2019 [Online].

\end{itemize}

\subsection{Download and Utilization}
\label{subsec:download}
The latest version of the toolbox can be downloaded directly from OSET at the address:
 \begin{center}
 \underline{\texttt{http://oset.ir/category.php?dir=Tools}}
  \end{center}
 
or from the public repository on GitHub, at:

 \begin{center} 
 \underline{\url{https://github.com/EsiSeraj/ERP_Connectivity_EMG_Analysis}}
 \end{center}

The functions and m-files can be downloaded separately as you need or all together in a compressed file named \texttt{ERP\_Connectivity\_EMG\_Analysis\_Toolbox}. Once you have downloaded and uncompressed the toolbox, 35 files represented in following Tables shall be appeared in your chosen directory. Table (1) represents all m-files (i.e. core functions, internal computational functions and demo test files) included in this toolset with their short descriptions and references. Additionally, a copy of both GNU general public license, this user manual and ten recorded sample data are also included (see Section~\ref{subsubsec:data} for details). Moreover, you might add the path of the directory in which you stored the toolbox to your MATLAB in order to easily use and apply the functions for your own dataset.


\begin{sidewaystable}[!hp]
\centering
\label{tab:AllOthermfiles}
\caption{Essential Motor Cortex Signal Analysis m-files}.
\begin{tabular}{ |c|c||c|c| }
\hline
row & name (function) & description & reference \\ 
 \hline
 \hline
 1 & \tt{test\_ERPanalysis\_main.m} & demo for testing the ERP analysis functions & Original \\ 
 \hline
 2 & \tt{test\_connectivity\_main.m} & demo for testing the functional connectivity functions & Original \\ 
 \hline
 3 & \tt{test\_EMGanalysis\_main.m} & demo for testing the EMG quantification functions & Original \\
 \hline 
 4 & \tt{trigger\_avg\_erp.m} & trigger-averaged ERP time-course estimation function & Original \\ 
 \hline
 5 & \tt{trigger\_avg\_TF\_erp.m} & trigger averaged ERP time/frequency representation & Original \\
 \hline
 6 & \tt{erp\_quantification.m} & quantifying the area of ERD/ERS events & Original \\
 \hline
 7 & \tt{TCPLV.m} & time-course PLV estimation function for a pair of electrodes & Original \\
 \hline
 8 & \tt{PWPLV.m} & pair-wise PLV estimation function & Original \\
 \hline
 9 & \tt{PWCoherence.m} & pair-wise coherence estimation function & Original \\
 \hline
 10 & \tt{PLV\_PhaseSeq.m} & PLV quantification function & OSET \cite{seraj2016cerebral} \\
 \hline
 11 & \tt{emg\_quantification.m} & quantifying the EMG signals & Original \\
 \hline
 \hline
 12 & \tt{phase\_est.m} & TFP phase estimation function  & OSET \cite{seraj2016cerebral} \\
 \hline
 13 & \tt{trigger\_synch.m} & synchronizing signals according to trigger time  & Original \\
 \hline
 14 & \tt{bdf2mat\_main.m} & reading EEG/EMG signal from bdf files, preprocessing and conditioning  & Original \\
 \hline
 15 & \tt{emg\_onset.m} & EMG onset estimation function  & Original \\
 \hline
 16 & \tt{drift\_reject.m} & drift cancellation from biological recordings  & OSET \cite{sameniopen} \\
 \hline
 17 & \tt{BaseLine2.m} & estimating the baseline of a signal  & OSET \cite{sameniopen} \\
 \hline
 18 & \tt{sig\_trend.m} & estimating the trend of a signal  & OSET \cite{sameniopen} \\
 \hline
 19 & \tt{BPFilter5.m} & forward-backward zero-phase CIC filtering function  & OSET \cite{sameniopen} \\
 \hline
 20 & \tt{task\_separator.m} & task based data separation to avoid memory overuse  & Original \\
 \hline
 21 & \tt{eeg\_read\_bdf.m} & Converting “*.bdf” to “*.mat”  & Gleb Tcheslavski \cite{GlebTcheslavski2007bdfreader} \\
 \hline
        
\end{tabular}
\end{sidewaystable}
%

\subsection{Getting Help}
\label{subsec:help}
If you have added the toolbox directory to the MATLAB\textregistered ~path you can simply type:
 \begin{center}
\texttt{<doc~~~function\_name>} \hspace*{1cm} or \hspace*{1cm} \texttt{<help~~~function\_name>}
 \end{center} 
 
in command window to get online help for the function you are using. Furthermore, you can also contact any of the authors\footnote{E. Seraj, K. Mahalingam: \texttt{\{eseraj3, kmahalingam\}@gatech.edu}} directly to ask any related questions or discuss possible difficulties or errors you might encounter. Please feel free to contact in either case.

\section{User Guide}
\label{sec:userguide}
\subsection{Overview}
\label{subsec:context}
This document is meant to help individuals use the "Essential Motor Cortex Signal Processing MATLAB Toolbox" which implements various methods for three major aspects of investigating human motor cortex from Neuroscience view point: (1) ERP estimation and quantification (e.g. based on \cite{pfurtscheller1999event, handy2005event, luck2014introduction, makeig2004mining}), (2) Cortical Functional Connectivity analysis (e.g. based on \cite{greicius2003functional, carter1973estimation, lachaux1999measuring, varela2001brainweb, rosenblum1996phase}) and (3) EMG quantification (e.g. based on \cite{ricamato2005quantification, walter1984temporal}). Measurements and methodologies are all derived based on the procedures suggested by the most popular reference articles in that category, as aforementioned.

The current version of toolbox covers six well-known and widely used approaches within Neuroscience community for analyzing motor cortex potentials and EMG signals, as follows:

\begin{itemize}
\item \textbf{Preconditioning:} Preprocessing the signals, including baseline drift removal, artifact rejection, movement onset detection (i.e. for non-cued movements), trigger synchronization (i.e. essential step for ERP extraction through non--synchronized trials of EEG data) and etc.

\item \textbf{Time-course ERP Dynamics:} Estimation and quantification of ERP (i.e. ERS and ERD events).

\item \textbf{Time-frequency ERP Analysis:} Estimation and representation of time-frequency (TF) maps through various widely-accepted TF analysis methods such as Short-time Fourier Transform (STFT), Continuous Wavelet Transform (CWT), Narrow-band Channelization (NBCH) and etc.

\item \textbf{Time-course Cortical Functional Connectivity:} Estimation and quantification of local/large -scale functional connectivity between arbitrary pairs of electrodes through Phase Locking Value (PLV) \cite{lachaux1999measuring, rosenblum1996phase} and Magnitude Squared Coherence (MSC) \cite{carter1973estimation}.

\item \textbf{Pair-wise Cortical Functional Connectivity:} Estimation and quantification of local/large -scale functional connectivity maps across scalp electrodes through PLV and MSC.

\item \textbf{EMG Quantification:} Estimation and quantification of electromyogram (EMG) records of active muscles during performing motor tasks.
\end{itemize} 

\subsection{Fundamentals}
\label{subsec:fundamentals}
\subsubsection{Data: Setup, Equipments and Format}
\label{subsubsec:data}
ActiveTwo is a very well-known commonly used high-resolution biosignal acquisition system that comes with advantageous capabilities such as active electrodes~\cite{biosemi2011biosemi}. Active electrodes are those that does not require extensive skin (scalp) preparation due to the presence of pre-amplifiers in them. The pin electrodes must be connected to the scalp on one end and to the acquisition system on the other. The acquisition system is then connected to the PC to which it relays electrode data serially. Electrode gel has to be used on the scalp with electrodes to improve conductivity and to reduce noise. Care should be taken to reduce noise as much as possible by properly grounding all electrical appliances connected with the electrode system, making sure there are minimum head movements while recording etc.

Before starting data collection, all electrodes need to be checked for their proper functionality. This can be performed through the single bucket test~\cite{smith2009activetwo}. Electrodes can be placed according to the 10-20 system~\cite{trans201210}, which is a very common approach for most studies. EEG cap has to be adjusted based on head circumference, nasion-inion, and ear-to-ear distance. Reference electrode areas are normally chosen to be the average between two ears~\cite{biosemi2011biosemi} and thus CMS/DRL flat type electrodes should be placed behind earlobes. Finally, EMG electrodes should be placed directly on the active muscle the activity of which is to be recorded.

The EEG signals recorded by ActiveTwo are stored in *.bdf format files. We provided a very useful MATLAB m-file by Gleb Tcheslavski~\cite{GlebTcheslavski2007bdfreader} that converts *.bdf format files to *.mat MATLAB data files.

\subsubsection{Preconditioning}
\label{subsec:preconditioning}
In ERP and functional connectivity analysis problems, preconditioning mostly refers to preprocessing the signals. This includes baseline drift removal, artifact rejection, movement onset detection from EMG signals, trigger synchronization (i.e. as described above) and etc.

\paragraph{Quick note on artifact rejection:} The concept of noise in this problem could be regarded to two different parts of recorded signals. First group are the common sources of noise observed in EEG recordings such as muscle activity, eye movement, electrocardiographic activity, instrumentation and equipment related artifacts and interference, slow drifts and amplifier saturation and etc. \cite{handy2005event}. Moreover, background EEG activity is considered as the other source of noise \cite{pfurtscheller1999event}. In this case, ERP is considered as signal and the background EEG activity as noise \cite{handy2005event, pfurtscheller1999event}. 

Artifact rejection in later case is normally easy, since as widely accepted, background ongoing EEG is considered as uncorrelated to signal (ERP) and thus, can be rejected by an ensemble averaging over trials \cite{pfurtscheller1999event, handy2005event, luck2014introduction}. The best way of dealing with the first group of noises however, is to not have any in first place. The easiest sources of noise to deal with are AC power lines, lighting and electronic equipment such as computers, displays and TVs, wireless routers, notebooks and also mobile phones. The basic step here is to simply remove any unnecessary sources of electro-magnetic (EM) noise from the recording room and, if possible, replace equipment using alternate current with direct current \cite{repovs2010dealing, light2010electroencephalography}. Further, a significant number of noise sources can be rejected through different steps of low-pass, high-pass and sometimes notch filters.

\begin{figure}[th]
\centering
\includegraphics[scale=0.75]{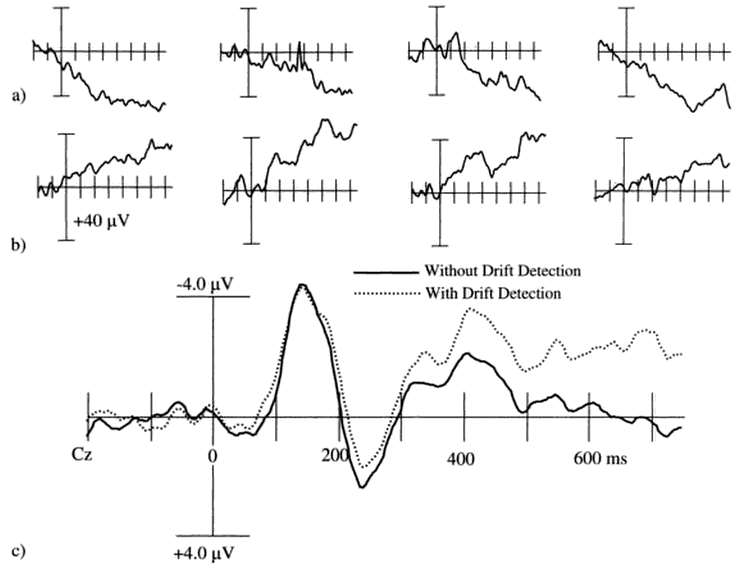}
\caption{The effect of drift (base line wander) on ERP analysis. (a) and (b) examples of trials with drifts. (c) estimated ERP in each case. This figure is adopted from \cite{handy2005event}.}
\label{fig:drifteffect}
\end{figure}

\paragraph{Drift cancellation:} Despite having various sources of additional noise on EEG, one of the most important ones to deal with, in this application, is the baseline drift noise (a.k.a base line wander). The drift noise normally happens due to sweating, drifts of cap and/or electrodes and similar reasons, which could normally lead to amplifier saturation \cite{handy2005event, repovs2010dealing} and thus incorrect and unreliable measurements. The effect of drifts on ERP estimation application is shown in Fig.~\ref{fig:drifteffect} which is directly employed from \cite{handy2005event}. As shown, although the effect of drift removal from 182 trials is relatively small, still is considerable for analyzing the relative decreases and increases in ERP power (ERD/ERS) quantification.

As suggested in \cite{handy2005event}, a number of issues are to be taken into consideration, such as detecting negative drifts as well as positive ones, using similar specs for the drift rejection method across trials and also choosing a proper-length (long enough) temporal window for drift calculation (preferably between 1 or two seconds) \cite{handy2005event}. Nevertheless, the filtering specs in EEG artifact removal are highly dependent on the data and recording conditions and thus, should be chosen empirically \cite{handy2005event, nitschke1998digital}. The strongest recommendation throughout the literature is to evaluate the filter before becoming committed to \cite{handy2005event, nitschke1998digital}. An important consideration to be taken into account here is that the filtering procedure itself, depending on the characteristics and nature of the utilized filter and filtering method, can also significantly affect the data in the non-filtered frequency ranges and thus can affect the onset detection and/or amplitude of estimated ERP. Therefore, to prevent these problems, it is highly recommended that filtering should be limited to what is necessary and unavoidable \cite{repovs2010dealing}.

\begin{figure}[th]
\centering
\includegraphics[scale=0.43]{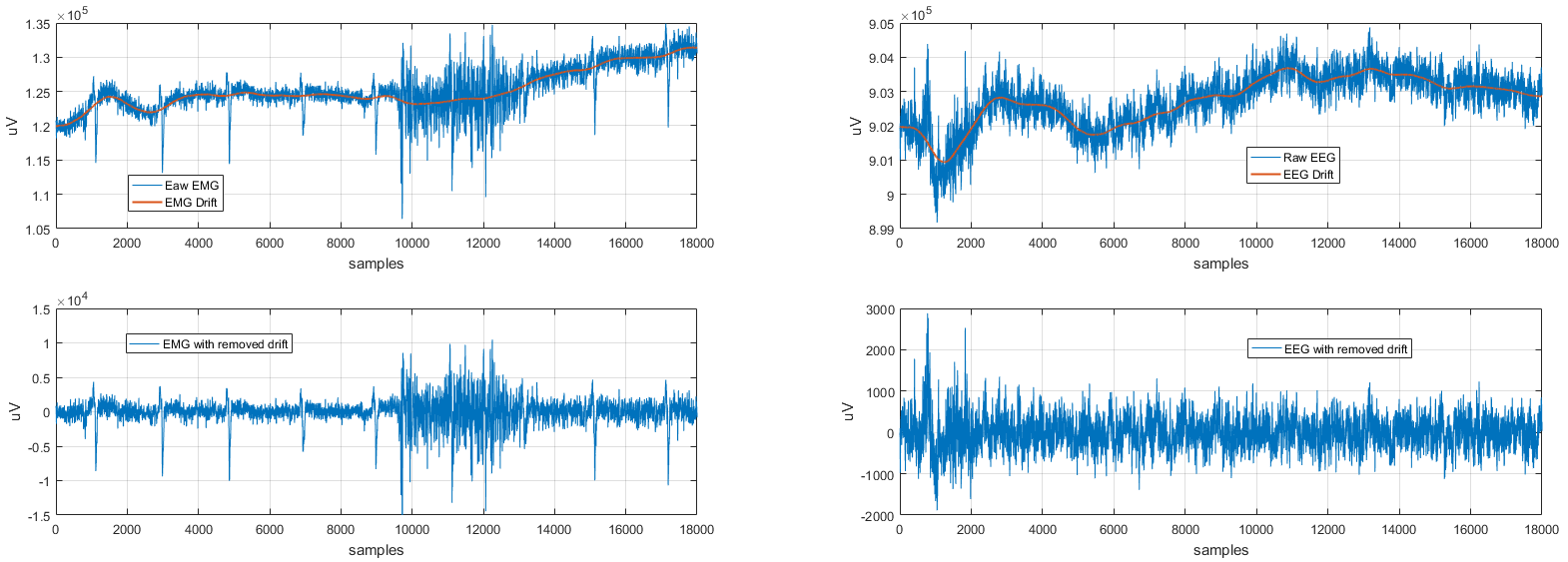}
\caption{Two sample EMG (left) and EEG (right) from provided sample dataset before and after removing the drift artifact as described in \cite{handy2005event, repovs2010dealing, nitschke1998digital}}
\label{fig:BLW_Removal}
\end{figure}

In this toolbox, the function \texttt{drift\_reject.m} is provided for this purpose which uses two stages of median or moving average temporal filtering to extract the base-line wander and reject it. To operate this function you can simply call function below with proper input parameters:
\begin{center}
\texttt{sig = drift\_reject(raw\_sig, L1, L2, approach)}
\end{center}
where the inputs and outputs are described later in Section \ref{sec:referencemanual}. Fig.~\ref{fig:BLW_Removal} is an elaboration of how how this function works.

\subsubsection{Time-course ERP Dynamics}
\label{subsubsec:timecourseERPdynamics}
For this part, the conventional approach, as proposed in \cite{pfurtscheller1999event}, is implemented. Accordingly, the well-known five-step method presented below is employed for ERP extraction from EEG signals:
\begin{itemize}
\item \textbf{Step \#1} – Band-pass frequency filtering the EEG data in each trial

\item \textbf{Step \#2} – Squaring the samples of filtered trials to find the power

\item \textbf{Step \#3} – Ensemble averaging over synchronized trials according to trigger time

\item \textbf{Step \#4} – Calculating the trend of output sequence in order to obtain ERP signal

\item \textbf{Step \#5} – Quantifying the estimated ERP signal according to a reference period
\end{itemize}

The estimated signal in Step \#4 is the desired ERP signal. Each step is elaborated below.

\paragraph{Step \#1:} For this step, zero-phase bandpass frequency filtering using Cascaded Integrator-Comb (CIC) filter is implemented \cite{lyons2005understanding}. The filter performs forward-backward filtering successively. It has absolutely zero-phase for \textit{even} filter order values and a phase-lag equal to a single stage CIC filter for \textit{odd} filter order values. Therefore, for this application we recommend an even value to be chosen as the filter order to prevent any subsequent phase distortion.
 
An important issue here is to select the frequency band of interest for showing the time course of ERP \cite{pfurtscheller1999event}. This concern is very crucial due to the fact that different brain states can modulate different frequency bands of EEG rhythms. Additionally, even for the same action, one might observe intra-trial variability on dominant frequencies \cite{pfurtscheller1999event}. Accordingly, estimating the time-course of ERP signal related to the task for further analysis requires additional cautions to be taken into account.
 
Due to the subjective process of choosing a proper frequency band for ERP analysis, herein, the functions are designed to accept frequency ranges in a way that users can customize their frequency range of interest and experiment various bands of frequency easily. To read further in these regards, refer to \cite{pfurtscheller1999event, handy2005event, luck2014introduction}.

\paragraph{Steps \#2, \#3 and \#4:} Steps 2 through 4 are closely related (e.g. even could be considered as one step) where after squaring the samples of the filtered signal, two successive \textit{averaging} steps are performed across trials and then time \cite{pfurtscheller1999event}. Ensemble averaging is the most important one among these steps since EEG signals are required to be averaged over synchronized data trials according to their respective trigger time. Based on this, first the onset of movement needs to be estimated (i.e. can be performed either through hardware settings (e.g. cue based recordings) or using our provided \texttt{emg\_onset.m} function) as the trigger time. Then, the EEG trials have to be sorted and aligned with respect to the trigger time and eventually averaged across data trials. Afterwards, a simple temporal averaging is performed on obtained signal to extract the ERP sequence.

Interested individuals can refer to either the Reference Manual (i.e. Section\ref{sec:referencemanual}) or \cite{pfurtscheller1999event, handy2005event, luck2014introduction} for details of each averaging step and implemented signal trend estimation function.

\paragraph{Steps \#5:} For this step, normally a 1-sec long reference period before the movement onset is chosen. Afterwards, the relative decrease intervals in power after movement onset are called ERD events and vice versa, the relative increase intervals in power are named ERS \cite{pfurtscheller1977graphical, graimann2006quantification}. 

According to discussions presented in \cite{pfurtscheller1977graphical, graimann2006quantification, pfurtscheller1999event}, reference period is chosen some seconds before the task onset, completely based on application of interest and data and thus is empirically set. Reference value is calculated as the average of power samples within reference period. An axis according to the reference value is set to 100\% and the corresponding crosses of ERP curve are named respectively, time step "0", start of ERD1/ERS1, end of ERD1/ERS1 and so on. Afterwards, standard deviation of the reference period is calculated and the corresponding confidence intervals are defined (mean plus/minus 2 or 3 times STD). Finally, the area enclosed by ERP curve and “\textbf{mean - confidence interval}” is calculated and normalized by the length of that segment as ERD area and vice versa; the area enclosed by ERP curve and “\textbf{mean + confidence interval}” is calculated and normalized by the length as ERS area. 

\begin{figure}[th]
\centering
\includegraphics[scale=0.65]{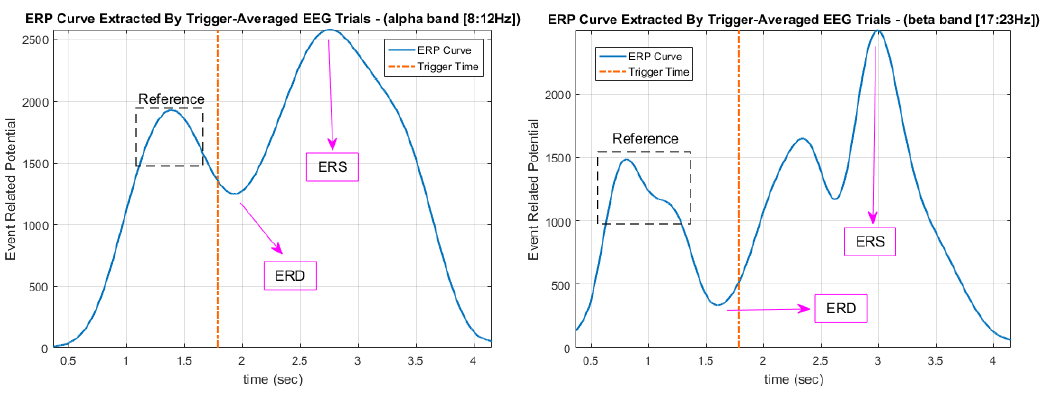}
\caption{Time courses of trigger averaged ERP curves, related to alpha and beta bands. }
\label{fig:TCERP}
\end{figure}

In this toolbox, the function \texttt{trigger\_avg\_erp.m} is provided for this purpose which utilizes \texttt{trigger\_ synch.m} function to synchronize cerebral signals according to their trigger time and \texttt{BPFilter5.m} function to perform the aforementioned CIC filtering. To operate this function you can use the demo m-file \texttt{test\_ERPanalysis\_main.m} or simply call function below with proper input parameters:
\begin{center}
\texttt{[erp, synch\_eeg, synch\_emg, trigger\_time\_sec, time\_vec] = trigger\_avg\_erp(eeg, emg, fs, emg\_onset\_sampl, freq\_band, duration)}
\end{center}
where the inputs and outputs are described later in Section \ref{sec:referencemanual}. Fig.~\ref{fig:TCERP} is an elaboration of how how this function works.

\begin{figure}[th]
\centering
\includegraphics[scale=0.57]{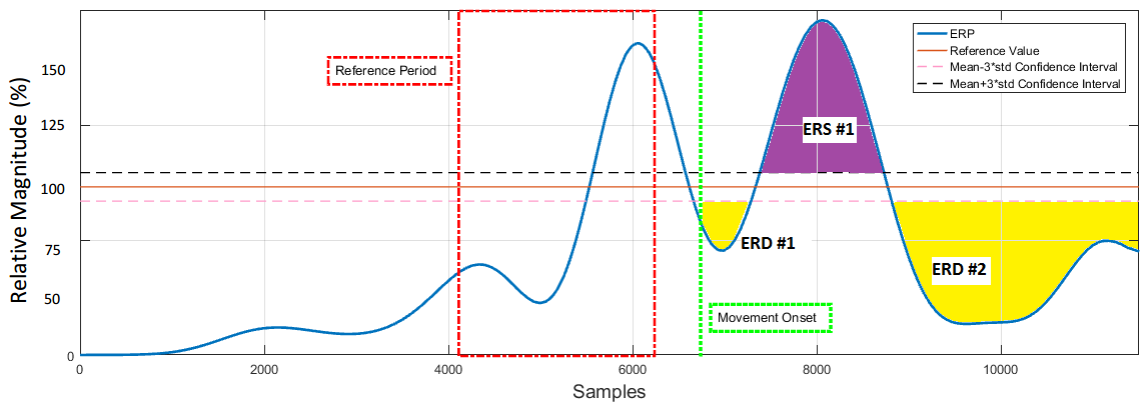}
\caption{Procedure for quantifying the area of ERD/ERS Events. Note that the confidence interval lines are plotted with exaggeration for better representation and are not in scale.}
\label{fig:ERP_area_quant}
\end{figure}

Moreover, function \texttt{erp\_quantification.m} is provided to quantify the calculated ERP signal as the output of \texttt{trigger\_avg\_erp.m} according to the exact procedure described above. To operate this function you can simply call function below with proper input parameters:
\begin{center}
\texttt{[erd\_area, ers\_area, quant\_erp] = erp\_quantification(erp, fs, trigger\_time\_sec, ref\_per, cof\_intv)}
\end{center}
where the inputs and outputs are described later in Section \ref{sec:referencemanual}. Fig.~\ref{fig:ERP_area_quant} is an elaboration of how how this function works.

\subsubsection{Time-frequency ERP Analysis}
\label{subsec:timefrequencyERPanalysis}
Time-frequency representations (TFRs) can monitor ERPs in a wide range of frequency components alongside with their temporal variations. Using TFRs, not only the risk of missing important information is lower, but also the interpretations provided can be more comprehensive and generic. Herein, as a part of ERP analysis package, a MATLAB function is dedicated to this purpose in which three different approaches for time/frequency analysis, namely Short-Time Fourier Transform (STFT), Continues Wavelet Transform (CWT) and Narrow-Band Channelization method (NBCH) as introduced in \cite{pfurtscheller1999event}, are implemented.
 
Theoretical concepts of each TFR method is out of scope of the current document, however, interested users can refer to \cite{cohen1995time} for a detailed and comprehensive discussion on time/frequency analysis concepts and its techniques. Nevertheless, as a short description, in STFT, the TF map of each trial is calculated separately through Fourier Transform and averaged over all. Similarly, for CWT the scalogram displaying the squared and over-all-trials-averaged wavelet coefficients are used. Eventually, for NBCH, ERP maps are generated according to ERP behaviors within very narrow frequency bands. The ERP maps are matrices, the rows of which correspond to frequency-specific ERP estimations.

\begin{figure}[th]
\centering
\includegraphics[scale=0.625]{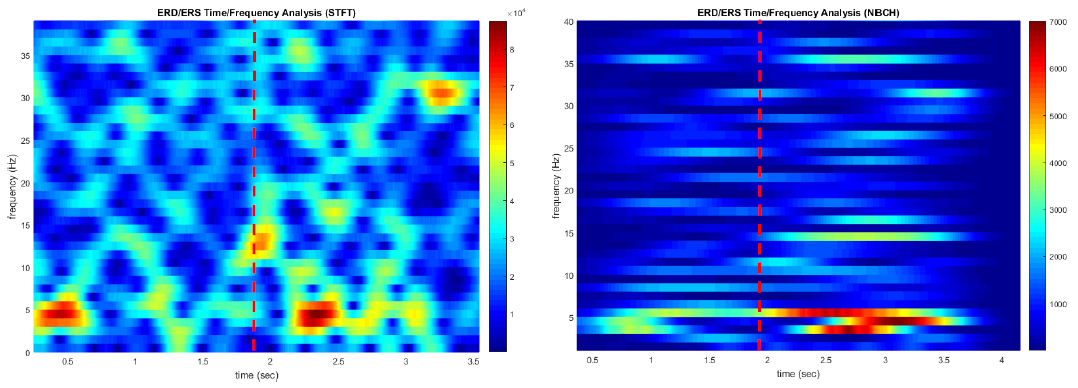}
\caption{TF maps estimated through STFT and NBCH methods. Note that despite the visible differences, both methods show almost same interpretations (note the relative magnitudes), with respect to the trigger time (red dashed line).}
\label{fig:TFERP}
\end{figure}

The implemented function enables variable frequency range. The procedures utilized for all three TFR methods here are in accordance with \cite{pfurtscheller1999event}. The function \texttt{trigger\_avg\_TF\_erp.m} is provided for this purpose which utilizes \texttt{trigger\_ synch.m} function to synchronize cerebral signals according to their trigger time and \texttt{BPFilter5.m} function to perform the aforementioned CIC filtering for NBCH method. Moreover, for STFT and CWT methods, MATLAB's inner \texttt{spectrogram} and \texttt{cwt} functions are used. To operate this function you can use the demo m-file \texttt{test\_ERPanalysis\_main.m} or simply call function below with proper input parameters:
\begin{center}
\texttt{[erp\_tf, synch\_eeg, synch\_emg, trigger\_time\_sec, freq\_vec, time\_vec] = trigger\_avg\_TF\_erp(eeg, emg, fs, emg\_onset\_sampl, duration, method)}
\end{center}
where the inputs and outputs are described later in Section \ref{sec:referencemanual}. Fig.~\ref{fig:TFERP} is an elaboration of how how this function works. It is worth noting that the CWT produces similar results. Note that several benchmark studies such as~\cite{seraj2017improved}, \cite{seraj2017investigation}, \cite{karimzadeh2018distributed} and \cite{seraj2019fmri} are research show-cases of the real-world applications how these TF analysis data can be put to use. For instance, as investigated and shown by~\cite{boostani2017comparative} and also \cite{karimzadeh2018distributed} and \cite{karimzadeh2015presenting} inspecting the irregularities of the TF patterns through entropy quantities can be very informative regarding the underlaying motor-task~\cite{seraj2017improved}, depth of sleep and Neurons activity rate~\cite{boostani2017comparative, karimzadeh2018distributed}, levels of Alzheimer’s disease~\cite{seraj2019instantaneous,seraj2019fmri} and even the cognitive skills and their respective performances in subjects~\cite{karimzadeh2017sleep, seraj2017investigation, seraj2016cerebral}.

\subsubsection{Time-course Cortical Functional Connectivity}
\label{subsubsec:timecourseCFC}
Time-course cortical functional connectivity is usually used to investigate the local-scale brain connectivity across different electrodes within same brain regions (i.e. primary motor cortex). To this end, Phase Locking Value (PLV) as the common quantity for this purpose \cite{lachaux1999measuring, rosenblum1996phase, seraj2016cerebralsynchrony} are implemented, measuring both temporal dynamics of local-scale and heat-maps of large-scale functional connectivity. The details of each method is described below.

\paragraph{Phase-locking Value (PLV):}
\label{parag:PLV}
PLV is a measure for quantifying how constant the phase difference between two signals is. In order to calculate the PLV in frequency $ f $ for two signals (or channels) $ x(t) $ and $ y(t) $, the following steps are required \cite{lachaux1999measuring, rosenblum1996phase}:
\begin{itemize}
\item Using narrow-band filters centered at $ f $, calculate the instantaneous frequency-specific phase values $ \phi_x(t, f) $ and $ \phi_y(t, f) $.
\item Calculate the instantaneous phase-difference between $ x(t) $ and $ y(t) $ and quantify the local stability of this phase-difference over time as follows:
\begin{equation}
\label{eq:plv}
PLV(f)=\left|\frac{1}{T}\sum_{t=1}^{T}\exp\left(j[\phi_y(t,f)-\phi_x(t,f)]\right)\right|
\end{equation}
where $ T $ is the signal length and the summation is over all temporal samples of the instantaneous phases.
\end{itemize}

PLV varies between 0 and 1, corresponding to completely non-synchronized signals and  complete synchronization, respectively \cite{lachaux1999measuring, rosenblum1996phase}. In simple wolds, considering the difference between firing rates of neurons as phase difference between their electrical potentials, PLV measures how "\textit{connected}" different neurons in far regions of brain are through quantifying the difference between their firing rates \cite{lachaux1999measuring, rosenblum1996phase, seraj2016cerebralsynchrony}.

One very important consideration is the phase estimation approach that one needs to take into account. For phase estimation in first step, the Transfer Function Perturbation (TFP) method introduced in \cite{sameni2017robust} and \cite{seraj2017robust} has been adopted. TFP is a statistical Monte-Carlo based phase estimation approach which uses infinitesimal perturbations on filter and other signal estimation parameters in order to account for stochastic properties of EEG signals. According to the discussion presented in Section~\ref{subsec:preconditioning}, using TFP is specifically required due to presence of background EEG noise (which in this case cannot be rejected as simple). It has been shown in \cite{sameni2017robust} and \cite{seraj2017robust} that without using TFP, estimated phase of EEG may contain variations and spikes which do not have any physiological origins within brain or are not due to brain activity and are merely side-effects of estimation method. For this purpose, cerebral signal phase estimation toolbox \cite{seraj2016cerebral} has been used.

\begin{figure}[th]
\centering
\includegraphics[scale=0.5]{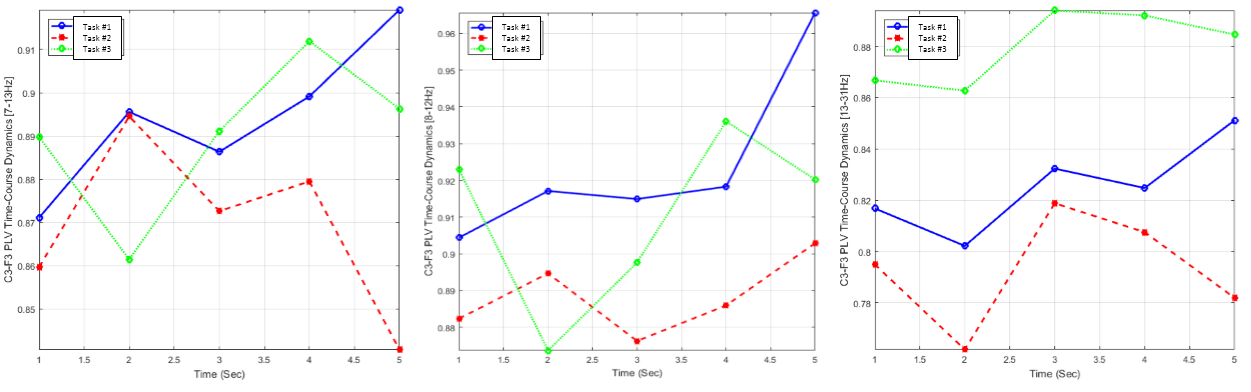}
\caption{C3-F3 PLV time-course dynamics for three different motor tasks in a clinical experiment measured for 7-13Hz, 8-12Hz and 13-30Hz rhythms from left to right respectively, within 1 second intervals. Movement onset is around 3.5sec.}
\label{fig:TCPLVs}
\end{figure}

\begin{figure}[th]
\centering
\includegraphics[scale=0.5]{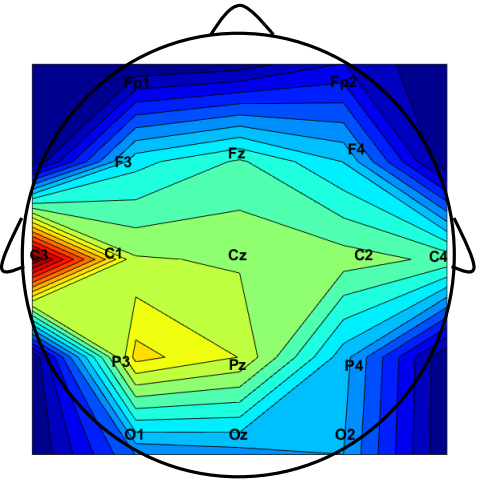}
\caption{Electrode placements for time-course PLV maps between C3 and all other EEG channels.}
\label{fig:plv_map_scalp_elec_with_head}
\end{figure}

\begin{figure}[th]
\centering
\includegraphics[scale=0.7]{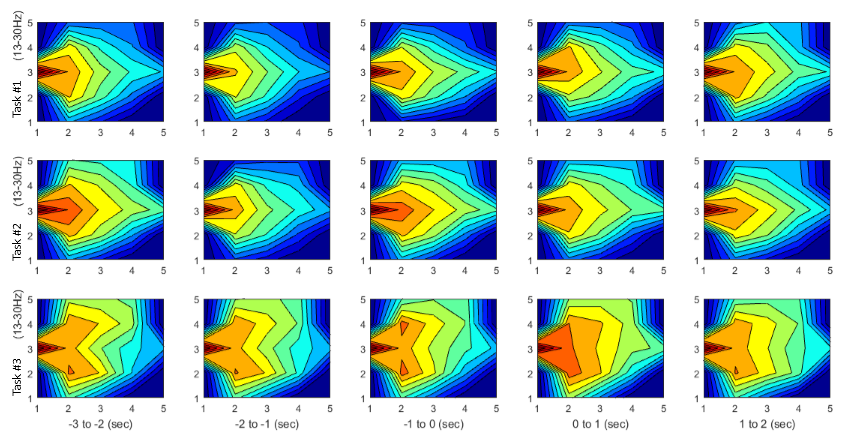}
\caption{Normalized time-course PLV maps between C3 and all other electrodes for beta rhythms (13-30Hz) for 1 second time bins from -3 prior to 2 seconds after the movement onset.}
\label{fig:plv_maps_normalized_beta}
\end{figure}

In this toolbox, function \texttt{TCPLV.m} is provided for this purpose which first utilizes \texttt{phase\_est.m} function to estimate the instantaneous phase (IP) sequence of input signals through introduced TFP method and then uses \texttt{PLV\_PhaseSeq.m} function to quantify the phase difference between calculated IPs. Function \texttt{TCPLV.m} measures time-course PLV for \textbf{\textit{1sec time-steps}} either between two specified electrode channels or between one important channel and all other channels (resulting in a heat-map across scalp). As recent examples of using such framework and implementation (time-course PLV) we can mention \cite{seraj2017robust, karimzadeh2018distributed, seraj2017improved}. To operate this function you can use the demo m-file \texttt{test\_connectivity\_main.m} or simply call function below with proper input parameters:
\begin{center}
\texttt{tcplv = TCPLV(eeg, fs, freq\_rng, pairofint, emg\_onset\_sampl, duration, pertnum)}
\end{center}
where the inputs and outputs are described later in Section \ref{sec:referencemanual}. Figures~\ref{fig:TCPLVs}, \ref{fig:plv_map_scalp_elec_with_head} and \ref{fig:plv_maps_normalized_beta} are elaborations of how this function works in either of aforementioned cases. Read the figure captions for more details.

\subsubsection{Pair-wise Cortical Functional Connectivity}
\label{subsubsec:pairwiseCFC}
Pair-wise cortical functional connectivity is usually used to investigate the large-scale brain connectivity across scalp (i.e. between different brain regions). To this end, the Magnitude Squared Coherence (MSC) and Phase Locking Value (PLV) as the common quantities for this purpose \cite{carter1973estimation, sun2004measuring, lachaux1999measuring, rosenblum1996phase, seraj2016cerebralsynchrony} are implemented, measuring heat-maps of large-scale functional connectivity.

\paragraph{Magnitude Squared Coherence (MSC):} The conventional method presented in \cite{carter1973estimation} for measuring coherence is based on weighted windowing of the Fourier transform of signals \cite{seraj2016cerebralsynchrony}. Considering $ x(t) $ and $ y(t) $ as two channels of recorded EEG signals, MSC can be measured in frequency $ f $ as below:
\begin{equation}
\label{eq:msc}
|MSC(f)|^2 = \frac{|PSD_{xy}(f)|^2}{PSD_{xx}(f)PSD_{yy}(f)} = \frac{|\sum_{i=1}^{N}X_i(f)Y_i^*(f)|^2}{\sum_{i=1}^{N}|X_i(f)|^2\sum_{i=1}^{N}|Y_i(f)|^2}
\end{equation}

The MSC here, is calculated using a non-overlapping hamming window using FFT analysis for EEG, in a frequency range of interest for all trials, and then averaged across trials. The mean MSC then is analyzed within \textbf{\textit{1 second temporal windows}} covering -3 to -2, -2 to -1, -1 to 0, 0 to 1 and 1 to 2 seconds, with respect to 0 set as trigger time. MSC measures are calculated for all possible electrode combinations, resulting in N$ \times $N MSC maps (for N-channel EEG recording), representing the functional connectivity across the whole scalp. Similar to PLV, MSC varies between 0 to 1 \cite{carter1973estimation, seraj2016cerebralsynchrony, gross2001dynamic}.

\begin{figure}[th]
\centering
\includegraphics[scale=0.6]{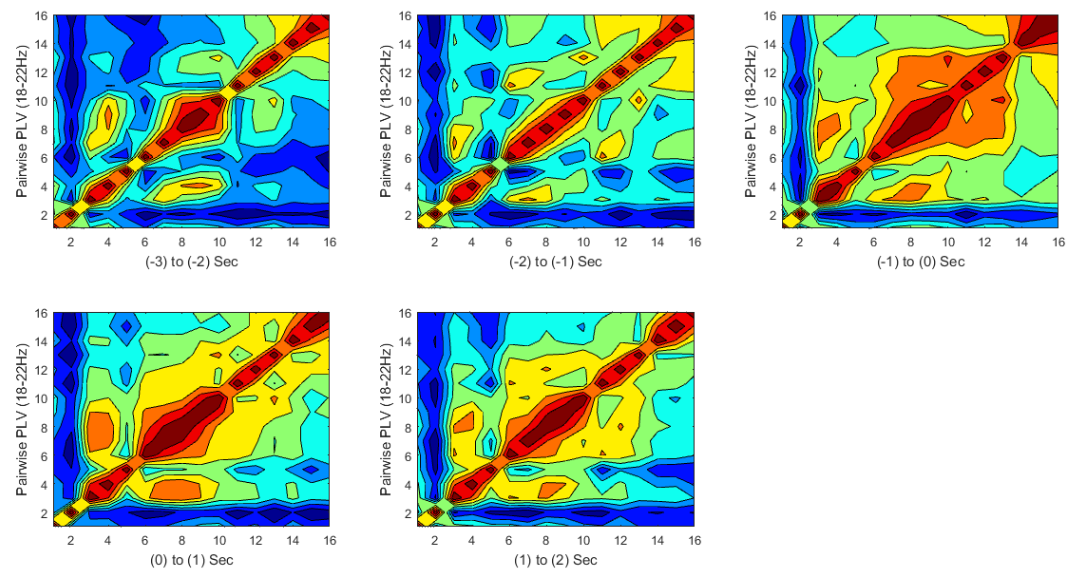}
\caption{MSC maps within 1 second time steps for an upper-body limb movement task.}
\label{fig:PWMSC}
\end{figure}

In this toolbox, function \texttt{PWCoherence.m} is provided for this purpose which first utilizes \texttt{BPFilter5.m} function to perform the aforementioned CIC filtering and then uses MATLAB's internal \texttt{mscohere.m} function to quantify the frequency-specific MSC values. As mentioned, function \texttt{PWCoherence.m} measures MSC maps for \textbf{\textit{1sec time-steps}} between all possible pairs of electrode. As recent examples of using such framework and implementation (pair-wise MSC) we can mention \cite{seraj2017investigation, karimzadeh2018distributed, seraj2017improved, seraj2017robust}. To operate this function you can use the demo m-file \texttt{test\_connectivity\_main.m} or simply call function below with proper input parameters:
\begin{center}
\texttt{pwcoher = PWCoherence(eeg, fs, freq\_rng, emg\_onset\_sampl, duration, plot\_flag)}
\end{center}
where the inputs and outputs are described later in Section \ref{sec:referencemanual}. Figures~\ref{fig:PWMSC} are elaborations of how how this function works in either of aforementioned cases. Read the figure captions for more details.

\paragraph{Phase Locking Value (PLV) for Large-scale Functional Connectivity:} Similar to MSC, PLV can be measured between all possible electrode pairs, resulting in PLV maps across entire scalp. We provided this option in this toolbox by implementing function \texttt{PWPLV.m} which first utilizes \texttt{phase\_est.m} function to estimate the instantaneous phase (IP) sequence of all channels of input data through introduced TFP method and then uses \texttt{PLV\_PhaseSeq.m} function to quantify the phase difference between calculated IPs. Function \texttt{TCPLV.m} measures pair-wise PLVs for \textbf{\textit{1sec time-steps}} between all possible electrode pairs resulting in N$ \times $N PLV maps (for N-channel EEG recording), representing the functional connectivity across entire scalp. As recent examples of using such framework and implementation (time-course PLV) we can mention \cite{sameni2017robust, seraj2017robust, karimzadeh2018distributed, seraj2017improved, seraj2017investigation}. To operate this function you can use the demo m-file \texttt{test\_connectivity\_main.m} or simply call function below with proper input parameters:
\begin{center}
\texttt{pwplv = PWPLV(eeg, fs, freq\_rng, emg\_onset\_sampl, duration, pertnum, plot\_flag)}
\end{center}
where the inputs and outputs are described later in Section \ref{sec:referencemanual}. This function also outputs figures similar to Figures~\ref{fig:PWMSC}.

\subsubsection{EMG Quantification} 
\label{subsubsec:emgquantification}
The purpose of this section of toolbox is to estimate and quantify electromyogram (EMG) signals recorded from active muscles during performing motor tasks. The EMG quantification is a common practice within Neuroscience community in order to evaluate the work-load of brain (through EMG curve's immediate slope after movement onset) and also for the rehabilitation purposes (stronger muscles after brain stimulation or training). The basic steps for EMG quantification are as below \cite{ricamato2005quantification}:
\begin{itemize}
\item Removing Electrocardiogram (ECG) artifacts from EMG signals of each trial
\item Form the full-rectified signal from the remaining EMG signal of each trial
\item Trigger-average the rectified signals across trials
\item Extract the trend of the averaged signal as the quantified EMG
\end{itemize}

The first step of this process, i.e. removing the ECG artifact, can be processed through several different simple or more complex approaches \cite{willigenburg2012removing}. High-pass frequency filters (HPF) \cite{drake2006elimination}, ECG subtraction through QRS-complex detection (FAS) \cite{aminian1988filtering}, adaptive filtering (AF) approaches \cite{abbaspour2014removing, nougarou2018efficient}, ICA based approaches \cite{abbaspour2015ecg, chen2016fastica} and also combined AF-ICA based methods \cite{li2013ecg}. Here in this toolbox however, we promise to stick with the simple and fast methods and thus, implement a modified high-pass frequency filtering approach not only to gain from the simplicity but to also improve the performance accuracy of removing ECG artifact from EMG signals. Accordingly, our modified ECG removal approach includes low-pass filter combined with median filtering (\textbf{LPF+MF}). HPF is normally used only to remove the major ECG frequency components which are restricted by the (1-30Hz) \cite{akselrod1981power}. The main drawback of this approach is due to the fact that HPF will remove the frequency components of EMG too \cite{akselrod1981power}. Nevertheless, to overcome this limitation, we implemented and used a LPF+MF to first detect the ECG signal (instead of blindly removing its components) and then perform a temporal subtraction. To this end, first we track the major ECG frequency components (1-30Hz) with a fourth order (can be modified within function parameters) elliptic IIR low-pass filter with pass-band ripple and stop-band attenuations of 0.1dB and -50dB, respectively. As before, the filtering is processed in a forward-backward zero-phase manner to avoid any phase distortion due to non-linear phase response of IIR filters. Afterwards, a median temporal filter with \textbf{\textit{small}} window-length is used to estimate and extract the ECG patterns more accurately \cite{ricamato2005quantification, akselrod1981power, barlaam2011time, krigolson2017event}.

\begin{figure}[th]
\centering
\includegraphics[scale=0.63]{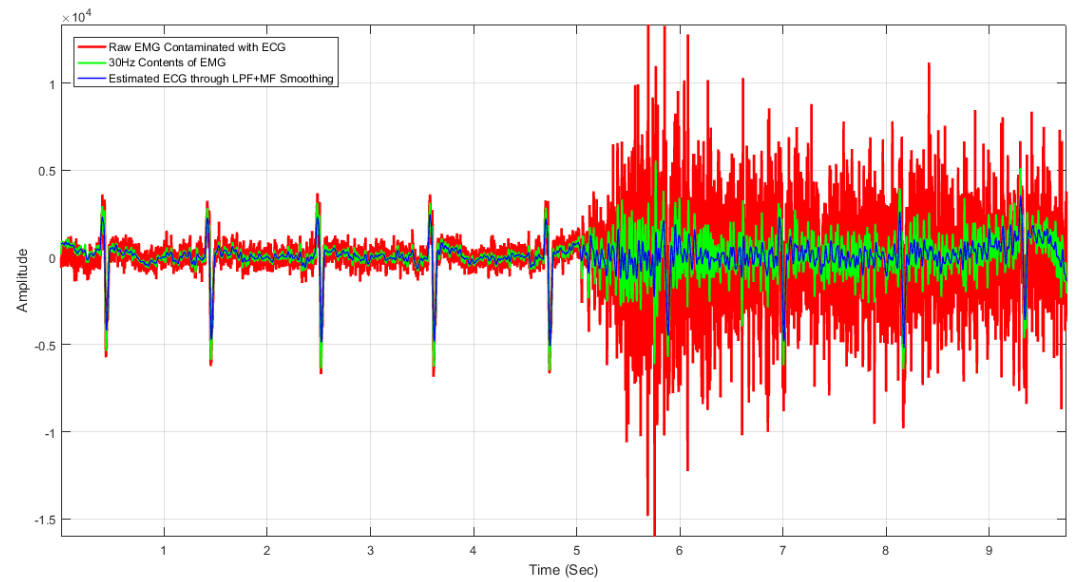}
\caption{The aforementioned procedure suggested to extract the ECG patterns from EMG signal. Note the final blue curve as the extracted ECG signal. The blue curve is then subtracted from the red one to form the ECG-free EMG signal.}
\label{fig:ECGRemoval}
\end{figure}

\begin{figure}[th]
\centering
\includegraphics[scale=0.61]{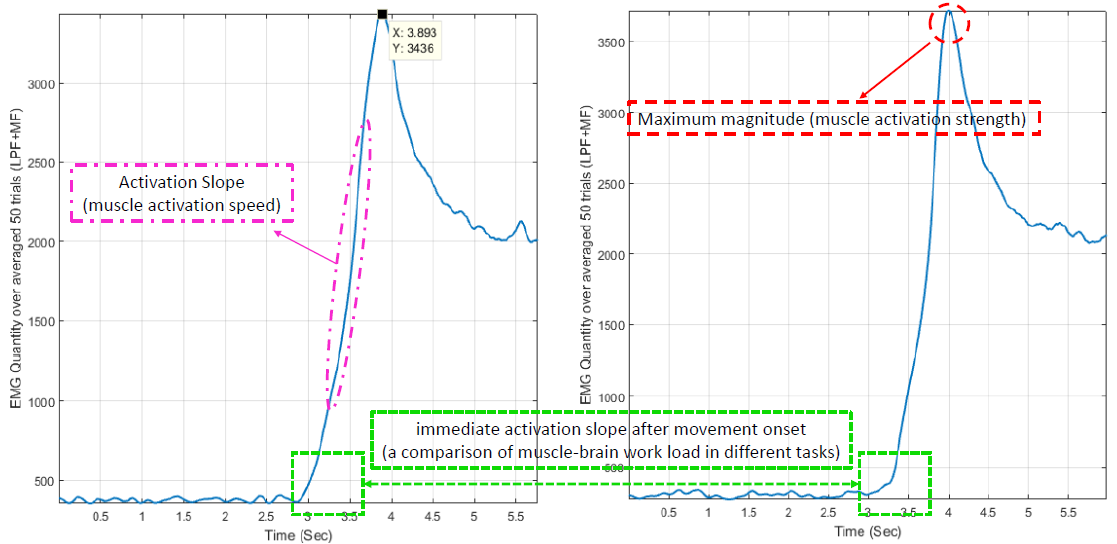}
\caption{Quantified EMG curves over 50 trials of two different tasks. Note the represented important parameters derived from quantified EMG.}
\label{fig:EMGparams}
\end{figure}

In this toolbox, function \texttt{emg\_quantification.m} is provided for this purpose which first utilizes MATLAB's internal \texttt{ellip} function to generate an elliptic IIR filter and then uses \texttt{BaseLine2.m} function twice; first to track and remove the ECG components and then to quantify the EMG signal as described before. To operate this function you can use the demo m-file \texttt{test\_EMGanalysis\_main.m} or simply call function below with proper input parameters:
\begin{center}
\texttt{[emg\_bl, synch\_emg2, ecg\_estimate2\_bl, time\_vec] = emg\_quantification(emg\_data, fs, emg\_onset\_sampl, duration)}
\end{center}
where the inputs and outputs are described later in Section \ref{sec:referencemanual}. Fig~\ref{fig:ECGRemoval} is an elaboration of how this function works. Performance of simple HPF is also shown for comparison and elaborating the reliability of the propose approach (i.e. LPF+MF). Maximum magnitude (muscle activation strength), activation slope (muscle activation speed) and immediate activation slope after movement onset (muscle-brain work load) are some potentially important information captured from quantified EMG through our provided function. See Fig~\ref{fig:EMGparams} for graphical elaboration of these concepts.

\paragraph{EMG Movement Onset Detection:}
\label{parag:emgonsetdetection}
As promised before in Section~\ref{subsubsec:timecourseERPdynamics}, here, we provide users with a simple but fast and reliable EMG onset estimation function. The onset detection in EMG signals has a broad literature in which various methods aimed to satisfy different purposes and applications are proposed and discussed. The simplest proposed approach for this purpose is the single thresholding method \cite{cavanagh1979electromechanical} which later on has been modified into double-threshold methods \cite{bonato1998statistical, drapala2012two}. Although there are several complex highly sensitive methods for onset detection, their utility really depends on users' application of interest and thus to reduce computational cost, here we designed and implemented a simple two-stage thresholding method which is highly reliable for ERP application. The proposed two-stage thresholding method is basically formed based on double-thresholding framework (which includes three steps) \cite{bonato1998statistical, drapala2012two}. The three steps are (1) signal conditioning (which is done in preprocessing phase), (2) detection of an event (performed by calculating a windowed standard deviation over time, named as STD vector) and (3) exact detection of the movement time (performed by using a threshold on STD vector's temporal trend). As reported previously in \cite{bonato1998statistical} and \cite{drapala2012two}, moving temporal filtering and thresholding methods are very common and could be satisfactory based on application requirements. Additionally, according to discussions in \cite{bonato1998statistical} and \cite{drapala2012two}, parameters and thresholds in such approaches need to be set empirically based on the data and application.

\begin{figure}[th]
\centering
\includegraphics[scale=0.8]{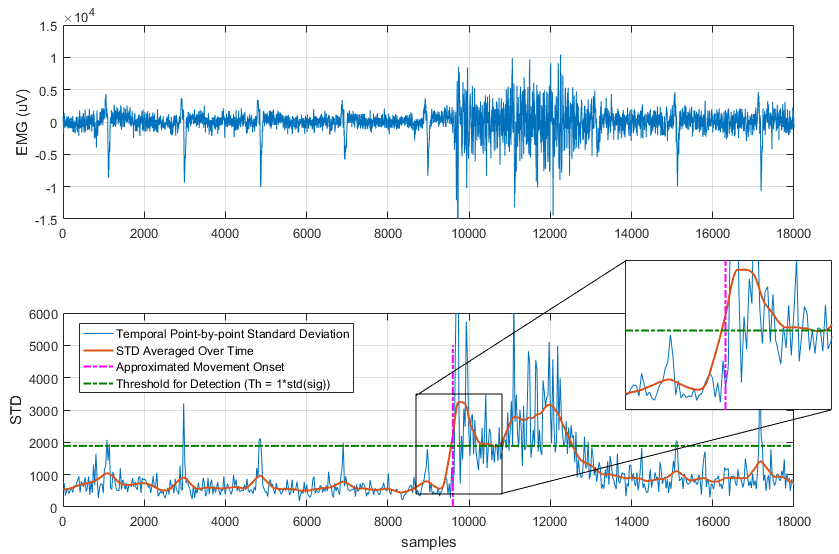}
\caption{The implemented EMG movement onset detection procedure.}
\label{fig:emgonsetdetection}
\end{figure}

In this toolbox, function \texttt{emg\_onset.m} is provided for this purpose which follows the exact procedure as described above. To operate this function you can use the demo m-file \texttt{test\_EMGanalysis\_main.m} or simply call function below with proper input parameters:
\begin{center}
\texttt{[onset\_sampl, onset\_time] = emg\_onset(emg, fs, W, th\_coeff, Trl)}
\end{center}
where the inputs and outputs are described later in Section \ref{sec:referencemanual}. Fig~\ref{fig:emgonsetdetection} is an elaboration of how this function works. It worth noting that, although the implemented procedure is not as sensitive as complex new methods, it still satisfies our requirements for the application of ERP analysis \cite{barlaam2011time, krigolson2017event}. This is due to the fact that ERP is calculated by averaging over a large number of trials  and thus here we are interested in \textbf{intra-trial} variations and information rather than \textbf{inter-trial} ones.

\section{Reference Manual}
\label{sec:referencemanual}

\subsection{\texttt{bdf2mat\_main.m}}
\subsection*{\fbox{\parbox{14.7cm}{\texttt{bdf2mat\_main.m}}}}
\paragraph*{Purpose:}
Reading EEG/EMG signal from bdf files, preprocessing and conditioning.
 
\paragraph*{Synopsis (global mode):}
\begin{center}
{\tt
 [eeg\_data, emg\_data, fs] = bdf2mat\_main(trl\_num, elec\_num, emg\_ch, eeg\_ch, filename) 
}
\end{center}

\paragraph*{Synopsis (local mode):}
\begin{center}
{\tt
 [eeg\_data, emg\_data, fs, emg\_onset\_sampl, emg\_onset\_time] = bdf2mat\_main(trl\_num, elec\_num, emg\_ch, eeg\_ch, filename, drift\_flag, onset\_flag)
}
\end{center}

\paragraph*{Inputs:}
\begin{center}
\begin{tabular}{ll}
Input              & Description \\
\hline
\hline
\texttt{trl\_num} & number of trials \\
\texttt{elec\_num} & number of electrodes used to record EEG \\
\texttt{emg\_ch} & EMG channel of interest \\
\texttt{eeg\_ch} & scalar or double vector of EEG channels of interest \\
\texttt{filename} & file name format as a string \\
\hline
\texttt{drift\_flag} & baseline drift rejection flag, (options:\texttt{'drift'}, \texttt{'nodrift'}) \\
\texttt{onset\_flag} & onset detection flag, (options: 1, 0)\\
\end{tabular}
\end{center}

\paragraph*{Defaults:}
\begin{center}
\begin{tabular}{lc}
Input              & Default Values \\
\hline
\hline
\texttt{drift\_flag} & \texttt{'drift'} \\
\texttt{onset\_flag} & 1 \\
\end{tabular}
\end{center}

\paragraph*{Outputs:}
\begin{center}
\begin{tabular}{ll}
Output              & Description \\
\hline
\hline
\texttt{eeg\_data} & preprocessed EEGs of the selected channel from all trials \\
\texttt{emg\_data} & preprocessed EMGs of the selected channel from all trials \\
\texttt{fs} & sampling frequency (Hz) \\
\hline
\texttt{emg\_onset\_sampl} & sample number of the movement onset \\
\texttt{emg\_onset\_time} & corresponding time of movement onset (Seconds) \\
\end{tabular}
\end{center}

\paragraph*{Notes:}
\begin{itemize}
\item An empty bracket [.] Must be assigned to not-specified values.

\item EMG and EEG channels of interest have to be a doubles (i.e. either an integer or a vector of indices).

\item \texttt{filename} needs to be defined in a way that \texttt{trl\_num} can be used as an index to track and load all of the stored data. For instance, data file names may be defined as \texttt{filename\_i} where $ i=1, 2, ..., N $ in which N represents the total number of data files. Refer to our sample data included in this toolbox to see another example of this, if needed.

\item One should expect the \texttt{emg\_onset\_sampl} and \texttt{emg\_onset\_time} within output arguments in case they flagged the onset detection as 1 within input.
\end{itemize}

\subsection{\texttt{trigger\_synch.m}}
\subsection*{\fbox{\parbox{14.7cm}{\texttt{trigger\_synch.m}}}}
\paragraph*{Purpose:}
Synchronizing EEG/EMG signals according to movement onset time.
 
\paragraph*{Synopsis (global mode):}
\begin{center}
{\tt
 [ensemble\_eeg, synch\_eeg, trigger\_time\_sec, time\_vec] = trigger\_synch(eeg, fs, onset\_time)
}
\end{center}

\paragraph*{Synopsis (local mode):}
\begin{center}
{\tt
 [ensemble\_eeg, synch\_eeg, trigger\_time\_sec, time\_vec, synch\_emg] = trigger\_synch(eeg, fs, onset\_time, duration, emg)
}
\end{center}

\paragraph*{Inputs:}
\begin{center}
\begin{tabular}{ll}
Input              & Description \\
\hline
\hline
\texttt{eeg} & cell array containing EEG channels of interest from all trials \\
\texttt{fs} & sampling frequency (Hz) \\
\texttt{onset\_time} & vector of onset times (Seconds) \\
\hline
\texttt{duration} & required signal duration after movement onset (Seconds) \\
\texttt{emg} & cell array containing EMG channels of interest from all trials\\
\end{tabular}
\end{center}

\paragraph*{Defaults:}
\begin{center}
\begin{tabular}{lc}
Input              & Default Values \\
\hline
\hline
\texttt{duration} & 2 \\
\texttt{emg} & \{.\}\\
\end{tabular}
\end{center}

\paragraph*{Outputs:}
\begin{center}
\begin{tabular}{ll}
Output              & Description \\
\hline
\hline
\texttt{ensemble\_eeg} & ensembles of EEG trials in a matrix \\
\texttt{synch\_eeg} & synchronized EEG signals based on trigger time \\
\texttt{trigger\_time\_sec} & trigger onset flag (Seconds) \\
\texttt{time\_vec} & time vector required for ERP plots \\
\hline
\texttt{synch\_emg} & synchronized EMG signals based on trigger time \\
\end{tabular}
\end{center}

\paragraph*{Notes:}
\begin{itemize}
\item An empty bracket [.] Must be assigned to not-specified values.

\item The provided function '\texttt{bdf2mat\_main.m}' outputs the compatible EEG/EMG data for this function; however, you can simply store your data in a cell array in which each element includes one data trial(s) of interest.

\item The output argument '\texttt{ensemble\_eeg}' is an essential variable to be used within the provided function '\texttt{trigger\_avg\_erp.m}' and thus you can simply replace it with \texttt{\~} symbol if using this function independently.

\item One should expect the \texttt{synch\_emg} (e.g. synchronized EMG data) within output arguments in case they inputed EMG data.
\end{itemize}

\subsection{\texttt{trigger\_avg\_erp.m}}
\subsection*{\fbox{\parbox{14.7cm}{\texttt{trigger\_avg\_erp.m}}}}
\paragraph*{Purpose:}
Trigger-averaged ERP time-course estimation function to extract the temporal dynamics of event-related potentials.
 
\paragraph*{Synopsis (global mode):}
\begin{center}
{\tt
 [erp, synch\_eeg, trigger\_time\_sec, time\_vec] = trigger\_avg\_erp(eeg, fs, freq\_band, onset\_time)
}
\end{center}

\paragraph*{Synopsis (local mode):}
\begin{center}
{\tt
 [erp, synch\_eeg, trigger\_time\_sec, time\_vec, synch\_emg] = trigger\_avg\_erp(eeg, fs, freq\_band, onset\_time, duration, emg)
}
\end{center}

\paragraph*{Inputs:}
\begin{center}
\begin{tabular}{ll}
Input              & Description \\
\hline
\hline
\texttt{eeg} & cell array containing EEG channels of interest from all trias \\
\texttt{fs} & sampling frequency (Hz)\\
\texttt{freq\_band} & frequency band of interest (refer to Notes below) \\
\texttt{onset\_time} & vector of onset times (Seconds) \\
\hline
\texttt{duration} & required signal duration after movement onset (Seconds) \\
\texttt{emg} & cell array containing EMG channels of interest from all trials \\
\end{tabular}
\end{center}

\paragraph*{Defaults:}
\begin{center}
\begin{tabular}{lc}
Input              & Default Values \\
\hline
\hline
\texttt{duration} & 2 \\
\texttt{emg} & \{.\} \\
\end{tabular}
\end{center}

\paragraph*{Outputs:}
\begin{center}
\begin{tabular}{ll}
Output              & Description \\
\hline
\hline
\texttt{erp} & extracted ERP signal \\
\texttt{synch\_eeg} & synchronized EEG signals based on trigger time \\
\texttt{trigger\_time\_sec} & trigger onset flag (Seconds) \\
\texttt{time\_vec} & time vector required for ERP plots \\
\hline
\texttt{synch\_emg} & synchronized EMG signals based on trigger time \\
\end{tabular}
\end{center}

\paragraph*{Notes:}
\begin{itemize}
\item An empty bracket [.] Must be assigned to not-specified values.

\item Available options for \texttt{freq\_band} are: '\texttt{delta}', '\texttt{theta}', '\texttt{alpha}', '\texttt{beta}' or '\texttt{gamma}' where are defined in ranges 1-4(Hz), 4-8(Hz), 8-12(Hz), 12-32(Hz) and 32-80Hz, respectively. In case you need to change these ranges, open up the function script and change \texttt{f0} and \texttt{bw} values for the band you wish to alter (i.e. lines 105 to 115). Note that \texttt{f0} and \texttt{bw} are center and bandwidth of the frequency range, respectively.

\item The provided function '\texttt{bdf2mat\_main.m}' outputs the compatible EEG/EMG data for this function; however, you can simply store your data in a cell array in which each element includes one data trial(s) of interest.

\item One should expect the \texttt{synch\_emg} (e.g. synchronized EMG data) within output arguments in case they inputed EMG data.
\end{itemize}

\subsection{\texttt{trigger\_avg\_TF\_erp.m}}
\subsection*{\fbox{\parbox{14.7cm}{\texttt{trigger\_avg\_TF\_erp.m}}}}
\paragraph*{Purpose:}
Trigger-averaged ERP T/F representation to track the time-frequency dynamics of event-related potentials.
 
\paragraph*{Synopsis (global mode):}
\begin{center}
{\tt
 [erp\_tf, synch\_eeg, trigger\_time\_sec, time\_vec, freq\_vec] = trigger\_avg\_TF\_erp(eeg, fs, onset\_time)
}
\end{center}

\paragraph*{Synopsis (local mode):}
\begin{center}
{\tt
 [erp\_tf, synch\_eeg, trigger\_time\_sec, time\_vec, freq\_vec, synch\_emg] = trigger\_avg\_TF\_erp(eeg, fs, onset\_time, duration, method, emg)
}
\end{center}

\paragraph*{Inputs:}
\begin{center}
\begin{tabular}{ll}
Input              & Description \\
\hline
\hline
\texttt{eeg} & cell array containing EEG channels of interest from all trials \\
\texttt{fs} & sampling frequency (Hz) \\
\texttt{onset\_time} & vector of onset times (Seconds)\\
\hline
\texttt{duration} & required signal duration after movement onset (Seconds) \\
\texttt{method} & method for T/F representation, (options: '\texttt{STFT}', '\texttt{CWT}', '\texttt{NBCH}') \\
\texttt{emg} & cell array containing EMG channels of interest from all trials\\
\end{tabular}
\end{center}

\paragraph*{Defaults:}
\begin{center}
\begin{tabular}{lc}
Input              & Default Values \\
\hline
\hline
\texttt{duration} & 2 \\
\texttt{method} & '\texttt{STFT}' \\
\texttt{emg} & \{.\}\\
\end{tabular}
\end{center}

\paragraph*{Outputs:}
\begin{center}
\begin{tabular}{ll}
Output              & Description \\
\hline
\hline
\texttt{erp\_tf} & estimated ERP time-frequency map \\
\texttt{synch\_eeg} & synchronized eeg signals based on trigger time \\
\texttt{trigger\_time\_sec} & trigger onset flag (Seconds) \\
\texttt{time\_vec} & time vector required for ERP plots \\
\texttt{freq\_vec} & frequency vector required for ERP map plots \\
\hline
\texttt{synch\_emg} & synchronized emg signals based on trigger time \\
\end{tabular}
\end{center}

\paragraph*{Notes:}
\begin{itemize}
\item An empty bracket [.] Must be assigned to not-specified values.

\item The provided function '\texttt{bdf2mat\_main.m}' outputs the compatible EEG/EMG data for this function; however, you can simply store your data in a cell array in which each element includes one data trial(s) of interest.

\item One should expect the \texttt{synch\_emg} (e.g. synchronized EMG data) within output arguments in case they inputed EMG data.
\end{itemize}

\subsection{\texttt{erp\_quantification.m}}
\subsection*{\fbox{\parbox{14.7cm}{\texttt{erp\_quantification.m}}}}
\paragraph*{Purpose:}
ERP area (ERD and ERS events' area) quantification.
 
\paragraph*{Synopsis (global mode):}
\begin{center}
{\tt
 [erd\_area, ers\_area, quant\_erp] = erp\_quantification(erp, fs, trigger\_time)
}
\end{center}

\paragraph*{Synopsis (local mode):}
\begin{center}
{\tt
 [erd\_area, ers\_area, quant\_erp] = erp\_quantification(erp, fs, trigger\_time, ref\_per, cof\_intv)
}
\end{center}

\paragraph*{Inputs:}
\begin{center}
\begin{tabular}{ll}
Input              & Description \\
\hline
\hline
\texttt{erp} & vector of estimated ERP signal \\
\texttt{fs} & sampling frequency (Hz) \\
\texttt{trigger\_time} & trigger onset flag (Seconds) \\
\hline
\texttt{ref\_per} & vector of reference period (refer to Notes below)\\
\texttt{cof\_intv} & confidence interval coefficient\\
\end{tabular}
\end{center}

\paragraph*{Defaults:}
\begin{center}
\begin{tabular}{lc}
Input              & Default Values \\
\hline
\hline
\texttt{ref\_per} & [-1.3, -0.3] Seconds \\
\texttt{cof\_intv} & 3 \\
\end{tabular}
\end{center}

\paragraph*{Outputs:}
\begin{center}
\begin{tabular}{ll}
Output              & Description \\
\hline
\hline
\texttt{erd\_area} & ERD events area \\
\texttt{ers\_area} & ERS events area \\
\texttt{quant\_erp} & ERP with quantified magnitude \\
\end{tabular}
\end{center}

\paragraph*{Notes:}
\begin{itemize}
\item An empty bracket [.] Must be assigned to not-specified values.

\item Reference period variable '\texttt{ref\_per}' have to be a double vector in [-a, -b] form where -a and -b are the edges of reference segment in Seconds. Minus sign shows that this period belongs to before movement onset (i.e. trigger time).

\item \texttt{erd\_area} and \texttt{ers\_area} are stored in a variable-size cell array the size of which depend on the duration of ERP signal after movement onset. For instance, in figure~\ref{fig:ERP_area_quant} there are 2 ERD events and only one ERS event detected. As a results, the corresponding area cell arrays will be of size 1$ \times $2 and 1$ \times $1, respectively.
\end{itemize}

\subsection{\texttt{TCPLV.m}}
\subsection*{\fbox{\parbox{14.7cm}{\texttt{TCPLV.m}}}}
\paragraph*{Purpose:}
PLV temporal dynamics estimated within 1 Second time-steps for any arbitrary time range before and after movement onset.
 
\paragraph*{Synopsis (global mode):}
\begin{center}
{\tt
 tcplv = TCPLV(eeg, fs, onset\_time)

}
\end{center}

\paragraph*{Synopsis (local mode):}
\begin{center}
{\tt
 tcplv = TCPLV(eeg, fs, onset\_time, freq\_rng, duration, pairofint, pertnum)
}
\end{center}

\paragraph*{Inputs:}
\begin{center}
\begin{tabular}{ll}
Input              & Description \\
\hline
\hline
\texttt{eeg} & cell array containing eeg channels of interest from all trials \\
\texttt{fs} & sampling frequency (Hz) \\
\texttt{onset\_time} & vector of onset times (Seconds)\\
\hline
\texttt{freq\_rng} & frequency range of interest (Hz) \\
\texttt{duration} & required temporal duration for tracking PLV dynamics (Seconds)\\
\texttt{pairofint} & channel pair of interest for PLV time-course\\
\texttt{pertnum} & number of perturbations while using the TFS phase estimation method \\
\end{tabular}
\end{center}

\paragraph*{Defaults:}
\begin{center}
\begin{tabular}{lc}
Input              & Default Values \\
\hline
\hline
\texttt{freq\_rng} & [12, 32] (Hz) \\
\texttt{duration} & [-3, 2] (Seconds) \\
\texttt{pairofint} & 'all' \\
\texttt{pertnum} & 100 \\
\end{tabular}
\end{center}

\paragraph*{Outputs:}
\begin{center}
\begin{tabular}{ll}
Output              & Description \\
\hline
\hline
\texttt{tcplv} & estimated time-course phase locking value (PLV) dynamics between eeg pairs of interest \\
\end{tabular}
\end{center}

\paragraph*{Notes:}
\begin{itemize}
\item An empty bracket [.] Must be assigned to not-specified values.

\item \texttt{freq\_rng} has to be in \texttt{[a, b]} form double vector where '\texttt{a}' and '\texttt{b}' are edges of the frequency band of interest in Hz.

\item \texttt{duration} has to be in [-a, b] form double vector where '\texttt{a}' is the required duration (Seconds) prior to movement onset and '\texttt{b}' is the required duration (Seconds) after the movement onset. The specified duration will be segmented into 1-sec-long windows to calculate PLV. Refer to Section~\ref{parag:PLV} for more details.

\item \texttt{pairofint} can either be a double vector or '\texttt{all}' string. As a double vector it has to be in [a, b] form resulting in PLV measures between electrode number \textbf{a} and electrode number \textbf{b}. Moreover, if you use '\texttt{all}', a PLV \textit{map} will be calculated between C3 electrode and all other electrodes to cover large-scale functional connectivity dynamics between motor cortex and other brain regions. In this case, C3 is chosen as the default electrode; however, if you wish to change this for your data, open up the script, scroll to line 143 of code and change the value 6 to you electrode number of interest.
\end{itemize}

\subsection{\texttt{PWPLV.m}}
\subsection*{\fbox{\parbox{14.7cm}{\texttt{PWPLV.m}}}}
\paragraph*{Purpose:}
Pair-wise Phase-Locking Value (PLV) dynamics estimated within 1-Second time-steps for any arbitrary time range and all electrode pairs.
 
\paragraph*{Synopsis (global mode):}
\begin{center}
{\tt
 pwplv = PWPLV(eeg, fs, onset\_time)

}
\end{center}

\paragraph*{Synopsis (local mode):}
\begin{center}
{\tt
 pwplv = PWPLV(eeg, fs, onset\_time, freq\_rng, duration, pertnum, plot\_flag)
}
\end{center}

\paragraph*{Inputs:}
\begin{center}
\begin{tabular}{ll}
Input              & Description \\
\hline
\hline
\texttt{eeg} & cell array containing eeg channels of interest from all trials \\
\texttt{fs} & sampling frequency (Hz) \\
\texttt{onset\_time} & vector of onset times (Seconds)\\
\hline
\texttt{freq\_rng} & frequency range of interest (Hz) \\
\texttt{duration} & required temporal duration for tracking PLV dynamics (Seconds)\\
\texttt{pertnum} & number of perturbations while using the TFS phase estimation method \\
\texttt{plot\_flag} & flag to visualize the results or not (options: '\texttt{plot}', '\texttt{noplot}')\\
\end{tabular}
\end{center}

\paragraph*{Defaults:}
\begin{center}
\begin{tabular}{lc}
Input              & Default Values \\
\hline
\hline
\texttt{freq\_rng} & [12, 32] (Hz) \\
\texttt{duration} & [-3, 2] (Seconds) \\
\texttt{pertnum} & 100 \\
\texttt{plot\_flag} & '\texttt{plot}' \\
\end{tabular}
\end{center}

\paragraph*{Outputs:}
\begin{center}
\begin{tabular}{ll}
Output              & Description \\
\hline
\hline
\texttt{pwplv} & estimated pairwise phase locking value (PLV) between all possible electrode pairs \\
\end{tabular}
\end{center}

\paragraph*{Notes:}
\begin{itemize}
\item An empty bracket [.] Must be assigned to not-specified values.

\item \texttt{freq\_rng} has to be in \texttt{[a, b]} form double vector where '\texttt{a}' and '\texttt{b}' are edges of the frequency band of interest in Hz.

\item \texttt{duration} has to be in [-a, b] form double vector where '\texttt{a}' is the required duration (Seconds) prior to movement onset and '\texttt{b}' is the required duration (Seconds) after the movement onset. The specified duration will be segmented into 1-sec-long windows to calculate PLV. Refer to Section~\ref{parag:PLV} for more details.

\item This function utilizes all of the electrodes (i.e. not just motor cortex electrode C3) and all possible combinations to calculate PLV values and generate connectivity maps. Refer to Section \ref{subsubsec:pairwiseCFC} for more details.
\end{itemize}

\subsection{\texttt{PWCoherence.m}}
\subsection*{\fbox{\parbox{14.7cm}{\texttt{PWCoherence.m}}}}
\paragraph*{Purpose:}
Pair-wise Magnitude-Squared Coherence (MSC) dynamics estimated within 1-Second time-steps for any arbitrary time range and all electrode pairs.
 
\paragraph*{Synopsis (global mode):}
\begin{center}
{\tt
 pwcoher = PWCoherence(eeg, fs, onset\_time)

}
\end{center}

\paragraph*{Synopsis (local mode):}
\begin{center}
{\tt
 pwcoher = PWCoherence(eeg, fs, onset\_time, freq\_rng, duration, plot\_flag)
}
\end{center}

\paragraph*{Inputs:}
\begin{center}
\begin{tabular}{ll}
Input              & Description \\
\hline
\hline
\texttt{eeg} & cell array containing eeg channels of interest from all trials \\
\texttt{fs} & sampling frequency (Hz) \\
\texttt{onset\_time} & vector of onset times (Seconds)\\
\hline
\texttt{freq\_rng} & frequency range of interest (Hz) \\
\texttt{duration} & required temporal duration for tracking PLV dynamics (Seconds)\\
\texttt{plot\_flag} & flag to visualize the results or not (options: '\texttt{plot}', '\texttt{noplot}')\\
\end{tabular}
\end{center}

\paragraph*{Defaults:}
\begin{center}
\begin{tabular}{lc}
Input              & Default Values \\
\hline
\hline
\texttt{freq\_rng} & [12, 32] (Hz) \\
\texttt{duration} & [-3, 2] (Seconds) \\
\texttt{plot\_flag} & '\texttt{plot}' \\
\end{tabular}
\end{center}

\paragraph*{Outputs:}
\begin{center}
\begin{tabular}{ll}
Output              & Description \\
\hline
\hline
\texttt{pwcoher} & estimated pairwise magnitude squared coherence (MSC) between all possible electrode pairs \\
\end{tabular}
\end{center}

\paragraph*{Notes:}
\begin{itemize}
\item An empty bracket [.] Must be assigned to not-specified values.

\item \texttt{freq\_rng} has to be in \texttt{[a, b]} form double vector where '\texttt{a}' and '\texttt{b}' are edges of the frequency band of interest in Hz.

\item \texttt{duration} has to be in [-a, b] form double vector where '\texttt{a}' is the required duration (Seconds) prior to movement onset and '\texttt{b}' is the required duration (Seconds) after the movement onset. The specified duration will be segmented into 1-sec-long windows to calculate PLV. Refer to Section~\ref{parag:PLV} for more details.

\item This function utilizes all of the electrodes (i.e. not just motor cortex electrode C3) and all possible combinations to calculate MSC values and generate connectivity maps. Refer to Section~\ref{subsubsec:pairwiseCFC} for more details.
\end{itemize}

\subsection{\texttt{emg\_quantification.m}}
\subsection*{\fbox{\parbox{14.7cm}{\texttt{emg\_quantification.m}}}}
\paragraph*{Purpose:}
EMG signal analysis and quantification.
 
\paragraph*{Synopsis (global mode):}
\begin{center}
{\tt
 [emg\_quant, synch\_emg2, ecg\_estimate2\_bl, time\_vec] = emg\_quantification(emg\_data, fs, emg\_onset\_sampl)
}
\end{center}

\paragraph*{Synopsis (local mode):}
\begin{center}
{\tt
 [emg\_quant, synch\_emg2, ecg\_estimate2\_bl, time\_vec] = emg\_quantification(emg\_data, fs, emg\_onset\_sampl, duration)
}
\end{center}

\paragraph*{Inputs:}
\begin{center}
\begin{tabular}{ll}
Input              & Description \\
\hline
\hline
\texttt{emg\_data} & cell array containing EMG signals of all trials \\
\texttt{fs} & sampling frequency (Hz) \\
\texttt{emg\_onset\_sampl} & vector containing movement onset samples of all trials \\
\hline
\texttt{duration} & duration of signal required after onset\\
\end{tabular}
\end{center}

\paragraph*{Defaults:}
\begin{center}
\begin{tabular}{lc}
Input              & Default Values \\
\hline
\hline
\texttt{duration} &2 \\
\end{tabular}
\end{center}

\paragraph*{Outputs:}
\begin{center}
\begin{tabular}{ll}
Output              & Description \\
\hline
\hline
\texttt{emg\_quant} & vector containing quantified EMG values \\
\texttt{synch\_emg2} & cell array containing synchronized EMG trials based on movement onset \\
\texttt{ecg\_estimate2\_bl} & extracted ECG signal from EMG channels \\
\texttt{time\_vec} & time-vector required for plotting quantified EMG \\
\end{tabular}
\end{center}


\subsection{\texttt{drift\_reject.m}}
\subsection*{\fbox{\parbox{14.7cm}{\texttt{drift\_reject.m}}}}
\paragraph*{Purpose:}
Drift (baseline wander) cancellation from biosignal recordings.
 
\paragraph*{Synopsis (global mode):}
\begin{center}
{\tt
sig = drift\_reject(raw\_sig, L1)
}
\end{center}

\paragraph*{Synopsis (local mode):}
\begin{center}
{\tt
sig = drift\_reject(raw\_sig, L1, L2, approach)
}
\end{center}

\paragraph*{Inputs:}
\begin{center}
\begin{tabular}{ll}
Input              & Description \\
\hline
\hline
\texttt{raw\_sig} & matrix or vector of raw recordings \\
\texttt{L1} & first stage window length (sample) \\
\hline
\texttt{L2} & second stage window length (sample) \\
\texttt{approach} & filtering approach, (options: 'md' or 'mn')\\
\end{tabular}
\end{center}

\paragraph*{Defaults:}
\begin{center}
\begin{tabular}{lc}
Input              & Default Values \\
\hline
\hline
\texttt{L2} & L1 (sample) \\
\texttt{approach} & 'mn'\\
\end{tabular}
\end{center}

\paragraph*{Outputs:}
\begin{center}
\begin{tabular}{ll}
Output              & Description \\
\hline
\hline
\texttt{sig} & matrix or vector of drift-rejected signals \\
\end{tabular}
\end{center}

\paragraph*{Notes:}
\begin{itemize}
\item An empty bracket [.] Must be assigned to not-specified values.
\end{itemize}

\subsection{\texttt{emg\_onset.m}}
\subsection*{\fbox{\parbox{14.7cm}{\texttt{emg\_onset.m}}}}
\paragraph*{Purpose:}
EMG onset detection function based on introduced two-stage thresholding approach in Section \ref{parag:emgonsetdetection}.
 
\paragraph*{Synopsis (global mode):}
\begin{center}
{\tt
[onset\_sampl, onset\_time] = emg\_onset(emg, fs, W) 
}
\end{center}

\paragraph*{Synopsis (local mode):}
\begin{center}
{\tt
 [onset\_sampl, onset\_time] = emg\_onset(emg, fs, W, th\_coeff, Trl)
}
\end{center}

\paragraph*{Inputs:}
\begin{center}
\begin{tabular}{ll}
Input              & Description \\
\hline
\hline
\texttt{emg} & the emg signal \\
\texttt{fs} & EMG signal's sampling frequency (Hz) \\
\texttt{W} & window length for STD calculation (sample) \\
\hline
\texttt{th\_coeff} & threshold coeff for onset detection \\
\texttt{Trl} & current trial number \\
\end{tabular}
\end{center}

\paragraph*{Defaults:}
\begin{center}
\begin{tabular}{lc}
Input              & Default Values \\
\hline
\hline
\texttt{th\_coeff} & 1 \\
\texttt{Trl} & --- \\
\end{tabular}
\end{center}

\paragraph*{Outputs:}
\begin{center}
\begin{tabular}{ll}
Output              & Description \\
\hline
\hline
\texttt{onset\_sampl} & sample number of the movement onset \\
\texttt{onset\_time} & corresponding time of movement onset \\
\end{tabular}
\end{center}

\paragraph*{Notes:}
\begin{itemize}
\item \texttt{th\_coeff} is a coefficient that will be multiplied by the standard deviation of the EMG baseline and by default is set to 1 so that the estimation threshold will be equal to one-standard-deviation of baseline (i.e. 1$ \times $STD\{emg\}).

\item An empty bracket [.] Must be assigned to not-specified values.
\end{itemize}

\subsection{\texttt{phase\_est.m}}
\subsection*{\fbox{\parbox{14.7cm}{\texttt{phase\_est.m}}}}
\paragraph*{Purpose:}
Instantaneous phase estimation by the Transfer Function Perturbation (TFP) method \cite{sameni2017robust, seraj2017robust, seraj2016cerebral}.
 
\paragraph*{Synopsis (global mode):}
\begin{center}
{\tt
 [phase\_avg, freq\_avg, amp\_avg, analytic\_sig\_avg] = phase\_est(sig, fs, f0, bw\_base)

}
\end{center}

\paragraph*{Synopsis (local mode):}
\begin{center}
{\tt
  [phase\_avg, freq\_avg, amp\_avg, analytic\_sig\_avg] = phase\_est(sig, fs, f0, bw\_base, pertnum)
}
\end{center}

\paragraph*{Inputs:}
\begin{center}
\begin{tabular}{ll}
Input              & Description \\
\hline
\hline
\texttt{sig} & input raw signal \\
\texttt{fs} & sampling frequency (Hz) \\
\texttt{f0} & center frequency of the passband (Hz) \\
\texttt{bw\_base} & bandwidth of the frequency filter (Hz) \\
\hline
\texttt{pertnum} & number of perturbations while using the TFP phase estimation method \\
\end{tabular}
\end{center}

\paragraph*{Defaults:}
\begin{center}
\begin{tabular}{lc}
Input              & Default Values \\
\hline
\hline
\texttt{pertnum} & 100 \\
\end{tabular}
\end{center}

\paragraph*{Outputs:}
\begin{center}
\begin{tabular}{ll}
Output              & Description \\
\hline
\hline
\texttt{phase\_avg} & estimated instantaneous phase of input signal \\
\texttt{freq\_avg} & estimated instantaneous frequency of input signal \\
\texttt{amp\_avg} & estimated instantaneous envelope of input signal \\
\texttt{analytic\_sig\_avg} & generated analytic form of input signal \\
\end{tabular}
\end{center}

\paragraph*{Notes:}
\begin{itemize}
\item \texttt{pertnum} is by default set to 100, however the value mostly depends on the application and also the required level of reliability. Nevertheless, 100 has been tested in several applications before and is considered enough \cite{seraj2017robust, karimzadeh2018distributed, seraj2017improved, seraj2017investigation}.

\item TFP perturbation parameters are set according to the findings of \cite{sameni2017robust} and are chosen in a way that are physiologically  irrelevant. In case one needs to change these values due to different application specs, refer to lines 83 to 88 of code.

\item An empty bracket [.] Must be assigned to not-specified values.
\end{itemize}

\subsection{\texttt{PLV\_PhaseSeq.m}}
\subsection*{\fbox{\parbox{14.7cm}{\texttt{PLV\_PhaseSeq.m}}}}
\paragraph*{Purpose:}
Calculating Phase Locking Value (PLV) matrix (Pairwise PLV) using phase sequences \cite{seraj2016cerebral}.
 
\paragraph*{Synopsis:}
\begin{center}
{\tt
 PLV = PLV\_PhaseSeq(phase\_sig)
 
 PLV = PLV\_PhaseSeq(phase\_sig1, phase\_sig2, phase\_sig3, ...)
}
\end{center}

\paragraph*{Inputs:}
\begin{center}
\begin{tabular}{ll}
Input              & Description \\
\hline
\hline
\texttt{phase\_sig} & input phase matrix \\
\hline
\texttt{phase\_sig1} & input phase vector \#1 \\
\texttt{phase\_sig2} & input phase vector \#2 \\
$ \bullet $ & $ \bullet $ \\
$ \bullet $ & $ \bullet $ \\
\end{tabular}
\end{center}

\paragraph*{Outputs:}
\begin{center}
\begin{tabular}{ll}
Output              & Description \\
\hline
\hline
\texttt{PLV} & Pairwise PLV matrix \\
\end{tabular}
\end{center}

\paragraph*{Notes:}
\begin{itemize}
\item While using the first case, the \texttt{phase\_sig} have to be a matrix with at least two rows where each row represents a phase signal. In second case, each of the \texttt{phase\_sig1...phase\_sign} are row vectors of phase sequences. This option is provided in case that someone needs to calculate the PLV matrix between separate phase signals.
\end{itemize}

\subsection{\texttt{sig\_trend.m}}
\subsection*{\fbox{\parbox{14.7cm}{\texttt{sig\_trend.m}}}}
\paragraph*{Purpose:}
Calculating the trend of a signal using local minima.
 
\paragraph*{Synopsis:}
\begin{center}
{\tt
 [Tr\_sig, loc] = sig\_trend(sig)
}
\end{center}

\paragraph*{Inputs:}
\begin{center}
\begin{tabular}{ll}
Input              & Description \\
\hline
\hline
\texttt{sig} & input raw signal \\

\end{tabular}
\end{center}

\paragraph*{Outputs:}
\begin{center}
\begin{tabular}{ll}
Output              & Description \\
\hline
\hline
\texttt{Tr\_sig} & trend vector \\
\texttt{loc} & returns the locations required for plotting the trend \\
\end{tabular}
\end{center}

\subsection{\texttt{task\_separator.m}}
\subsection*{\fbox{\parbox{14.7cm}{\texttt{task\_separator.m}}}}
\paragraph*{Purpose:}
Task based separation of data to avoid memory overuse.
 
\paragraph*{Synopsis (global mode):}
\begin{center}
{\tt
 sep\_file\_names = task\_separator(filename)
}
\end{center}

\paragraph*{Inputs:}
\begin{center}
\begin{tabular}{ll}
Input              & Description \\
\hline
\hline
\texttt{filename} & original whole-data file name as a string \\
\end{tabular}
\end{center}

\paragraph*{Outputs:}
\begin{center}
\begin{tabular}{ll}
Output              & Description \\
\hline
\hline
\texttt{sep\_file\_names} & cell array containing separated files' names \\
\end{tabular}
\end{center}

\section{Acknowledgment}
The authors would like to thank Dr. Maysam Ghovanloo\footnote{GT-Bionics Lab, School of Electrical and Computer Engineering, Georgia Tech, Atlanta, GA, USA}, Dr. Minoru Shinohara\footnote{Human Neuromuscular Physiology Lab, School of Biological Sciences, Georgia Tech, Atlanta, GA, USA}, Dr. Boris I. Prilutsky\footnote{Biomechanics and Motor Control Lab, School of Biological Sciences, Georgia Tech, Atlanta, GA, USA}, Dr. Andrew J. Butler\footnote{B.F. Lewis School of Nursing \& Health Professions, Georgia State University, Atlanta, GA, USA} and Zhenxuan Zhang\footnote{GT-Bionics Lab, School of Electrical and Computer Engineering, Georgia Tech, Atlanta, GA, USA} for their insightful discussions and comments.

\section{References}
 \bibliographystyle{IEEEtran}
 \bibliography{Refs_ERP_Manual}

\begin{thebibliography}{10}
\providecommand{\url}[1]{#1}
\csname url@samestyle\endcsname
\providecommand{\newblock}{\relax}
\providecommand{\bibinfo}[2]{#2}
\providecommand{\BIBentrySTDinterwordspacing}{\spaceskip=0pt\relax}
\providecommand{\BIBentryALTinterwordstretchfactor}{4}
\providecommand{\BIBentryALTinterwordspacing}{\spaceskip=\fontdimen2\font plus
\BIBentryALTinterwordstretchfactor\fontdimen3\font minus
  \fontdimen4\font\relax}
\providecommand{\BIBforeignlanguage}[2]{{%
\expandafter\ifx\csname l@#1\endcsname\relax
\typeout{** WARNING: IEEEtran.bst: No hyphenation pattern has been}%
\typeout{** loaded for the language `#1'. Using the pattern for}%
\typeout{** the default language instead.}%
\else
\language=\csname l@#1\endcsname
\fi
#2}}
\providecommand{\BIBdecl}{\relax}
\BIBdecl

\bibitem{pfurtscheller1999event}
G.~Pfurtscheller and F.~L. Da~Silva, ``Event-related eeg/meg synchronization
  and desynchronization: basic principles,'' \emph{Clinical neurophysiology},
  vol. 110, no.~11, pp. 1842--1857, 1999.

\bibitem{handy2005event}
T.~C. Handy, \emph{Event-related potentials: A methods handbook}.\hskip 1em
  plus 0.5em minus 0.4em\relax MIT press, 2005.

\bibitem{luck2014introduction}
S.~J. Luck, \emph{An introduction to the event-related potential
  technique}.\hskip 1em plus 0.5em minus 0.4em\relax MIT press, 2014.

\bibitem{makeig2004mining}
S.~Makeig, S.~Debener, J.~Onton, and A.~Delorme, ``Mining event-related brain
  dynamics,'' \emph{Trends in cognitive sciences}, vol.~8, no.~5, pp. 204--210,
  2004.

\bibitem{greicius2003functional}
M.~D. Greicius, B.~Krasnow, A.~L. Reiss, and V.~Menon, ``Functional
  connectivity in the resting brain: a network analysis of the default mode
  hypothesis,'' \emph{Proceedings of the National Academy of Sciences}, vol.
  100, no.~1, pp. 253--258, 2003.

\bibitem{carter1973estimation}
G.~Carter, C.~Knapp, and A.~Nuttall, ``Estimation of the magnitude-squared
  coherence function via overlapped fast fourier transform processing,''
  \emph{IEEE transactions on audio and electroacoustics}, vol.~21, no.~4, pp.
  337--344, 1973.

\bibitem{lachaux1999measuring}
J.-P. Lachaux, E.~Rodriguez, J.~Martinerie, and F.~J. Varela, ``Measuring phase
  synchrony in brain signals,'' \emph{Human brain mapping}, vol.~8, no.~4, pp.
  194--208, 1999.

\bibitem{varela2001brainweb}
F.~Varela, J.-P. Lachaux, E.~Rodriguez, and J.~Martinerie, ``The brainweb:
  phase synchronization and large-scale integration,'' \emph{Nature reviews
  neuroscience}, vol.~2, no.~4, p. 229, 2001.

\bibitem{rosenblum1996phase}
M.~G. Rosenblum, A.~S. Pikovsky, and J.~Kurths, ``Phase synchronization of
  chaotic oscillators,'' \emph{Physical review letters}, vol.~76, no.~11, p.
  1804, 1996.

\bibitem{ricamato2005quantification}
A.~L. Ricamato and J.~M. Hidler, ``Quantification of the dynamic properties of
  emg patterns during gait,'' \emph{Journal of electromyography and
  kinesiology}, vol.~15, no.~4, pp. 384--392, 2005.

\bibitem{walter1984temporal}
C.~B. Walter, ``Temporal quantification of electromyography with reference to
  motor control research,'' \emph{Human Movement Science}, vol.~3, no. 1-2, pp.
  155--162, 1984.

\bibitem{sameniopen}
R.~Sameni, ``The open-source electrophysiological toolbox (oset), version 3.1,
  2014,'' \emph{URL http://www. oset. ir}, 2014.

\bibitem{seraj2016cerebral}
E.~Seraj, ``Cerebral signal phase analysis toolbox--user guide,'' \emph{arXiv
  preprint arXiv:1610.02249}, 2016.

\bibitem{GlebTcheslavski2007bdfreader}
\BIBentryALTinterwordspacing
G.~Tcheslavski, ``eeg bdf reader matlab function.'' [Online]. Available:
  \url{https://www.mathworks.com/matlabcentral/fileexchange/13070-eeg-bdf-reader?focused=5083421&tab=function}
\BIBentrySTDinterwordspacing

\bibitem{biosemi2011biosemi}
B.~BioSemi, ``Biosemi activetwo.[eeg system],'' \emph{Amsterdam: BioSemi},
  2011.

\bibitem{smith2009activetwo}
L.~Smith, ``Activetwo system operating guidelines,'' \emph{Cortech Solutions,
  Inc.}, 2009.

\bibitem{trans201210}
T.~C. Technologies, ``10/20 system positioning manual.'' 2012.

\bibitem{repovs2010dealing}
G.~Repovs, ``Dealing with noise in eeg recording and data analysis,'' in
  \emph{Informatica Medica Slovenica}, vol.~15, no.~1, 2010, pp. 18--25.

\bibitem{light2010electroencephalography}
G.~A. Light, L.~E. Williams, F.~Minow, J.~Sprock, A.~Rissling, R.~Sharp, N.~R.
  Swerdlow, and D.~L. Braff, ``Electroencephalography (eeg) and event-related
  potentials (erps) with human participants,'' \emph{Current protocols in
  neuroscience}, vol.~52, no.~1, pp. 6--25, 2010.

\bibitem{nitschke1998digital}
J.~B. Nitschke, G.~A. Miller, and E.~W. Cook, ``Digital filtering in eeg/erp
  analysis: Some technical and empirical comparisons,'' \emph{Behavior Research
  Methods, Instruments, \& Computers}, vol.~30, no.~1, pp. 54--67, 1998.

\bibitem{lyons2005understanding}
R.~Lyons, ``Understanding cascaded integrator-comb filters,'' \emph{Embed Syst
  Program}, vol.~18, no.~4, pp. 14--27, 2005.

\bibitem{pfurtscheller1977graphical}
G.~Pfurtscheller, ``Graphical display and statistical evaluation of
  event-related desynchronization (erd),'' \emph{Electroencephalography and
  clinical neurophysiology}, vol.~43, no.~5, pp. 757--760, 1977.

\bibitem{graimann2006quantification}
B.~Graimann and G.~Pfurtscheller, ``Quantification and visualization of
  event-related changes in oscillatory brain activity in the time--frequency
  domain,'' \emph{Progress in brain research}, vol. 159, pp. 79--97, 2006.

\bibitem{cohen1995time}
L.~Cohen, \emph{Time-frequency analysis}.\hskip 1em plus 0.5em minus
  0.4em\relax Prentice hall, 1995, vol. 778.

\bibitem{seraj2017improved}
E.~Seraj and F.~Karimzadeh, ``Improved detection rate in motor imagery based
  bci systems using combination of robust analytic phase and envelope
  features,'' in \emph{Electrical Engineering (ICEE), 2017 Iranian Conference
  on}.\hskip 1em plus 0.5em minus 0.4em\relax IEEE, 2017, pp. 24--28.

\bibitem{seraj2017investigation}
E.~Seraj, ``An investigation on the utility and reliability of
  electroencephalogram phase signal upon interpreting cognitive responses in
  the brain: A critical discussion,'' \emph{Journal of Advanced Medical
  Sciences and Applied Technologies}, vol.~2, no.~4, pp. 299--312, 2017.

\bibitem{karimzadeh2018distributed}
F.~Karimzadeh, R.~Boostani, E.~Seraj, and R.~Sameni, ``A distributed
  classification procedure for automatic sleep stage scoring based on
  instantaneous electroencephalogram phase and envelope features,'' \emph{IEEE
  Transactions on Neural Systems and Rehabilitation Engineering}, vol.~26,
  no.~2, pp. 362--370, 2018.

\bibitem{seraj2019fmri}
E.~Seraj, M.~Yazdi, and N.~Shahparian, ``fmri based cerebral instantaneous
  parameters for automatic alzheimer's, mild cognitive impairment and healthy
  subject classification,'' \emph{arXiv preprint arXiv:1904.07441}, 2019.

\bibitem{boostani2017comparative}
R.~Boostani, F.~Karimzadeh, and M.~Nami, ``A comparative review on sleep stage
  classification methods in patients and healthy individuals,'' \emph{Computer
  methods and programs in biomedicine}, vol. 140, pp. 77--91, 2017.

\bibitem{karimzadeh2015presenting}
F.~Karimzadeh, E.~Seraj, R.~Boostani, and M.~Torabi-Nami, ``Presenting
  efficient features for automatic cap detection in sleep eeg signals,'' in
  \emph{2015 38th International Conference on Telecommunications and Signal
  Processing (TSP)}.\hskip 1em plus 0.5em minus 0.4em\relax IEEE, 2015, pp.
  448--452.

\bibitem{seraj2019instantaneous}
E.~Seraj, M.~Yazdi, and N.~Shahparian, ``Instantaneous fmri based cerebral
  parameters for automatic alzheimer, mild cognitive impairment and healthy
  subject classification,'' \emph{Journal of integrative neuroscience},
  vol.~18, no.~3, pp. 261--268, 2019.

\bibitem{karimzadeh2017sleep}
F.~Karimzadeh, M.~Nami, and R.~Boostani, ``Sleep microstructure dynamics and
  neurocognitive performance in obstructive sleep apnea syndrome patients,''
  \emph{Journal of integrative neuroscience}, vol.~16, no.~2, pp. 127--142,
  2017.

\bibitem{seraj2016cerebralsynchrony}
E.~Seraj, ``Cerebral synchrony assessment: A general review on cerebral
  signals' synchronization estimation concepts and methods,'' \emph{arXiv
  preprint arXiv:1612.04295}, 2016.

\bibitem{sameni2017robust}
R.~Sameni and E.~Seraj, ``A robust statistical framework for instantaneous
  electroencephalogram phase and frequency estimation and analysis,''
  \emph{Physiological measurement}, vol.~38, no.~12, p. 2141, 2017.

\bibitem{seraj2017robust}
E.~Seraj and R.~Sameni, ``Robust electroencephalogram phase estimation with
  applications in brain-computer interface systems,'' \emph{Physiological
  measurement}, vol.~38, no.~3, p. 501, 2017.

\bibitem{sun2004measuring}
F.~T. Sun, L.~M. Miller, and M.~D'esposito, ``Measuring interregional
  functional connectivity using coherence and partial coherence analyses of
  fmri data,'' \emph{Neuroimage}, vol.~21, no.~2, pp. 647--658, 2004.

\bibitem{gross2001dynamic}
J.~Gro{\ss}, J.~Kujala, M.~H{\"a}m{\"a}l{\"a}inen, L.~Timmermann,
  A.~Schnitzler, and R.~Salmelin, ``Dynamic imaging of coherent sources:
  studying neural interactions in the human brain,'' \emph{Proceedings of the
  National Academy of Sciences}, vol.~98, no.~2, pp. 694--699, 2001.

\bibitem{willigenburg2012removing}
N.~W. Willigenburg, A.~Daffertshofer, I.~Kingma, and J.~H. van Die{\"e}n,
  ``Removing ecg contamination from emg recordings: A comparison of ica-based
  and other filtering procedures,'' \emph{Journal of electromyography and
  kinesiology}, vol.~22, no.~3, pp. 485--493, 2012.

\bibitem{drake2006elimination}
J.~D. Drake and J.~P. Callaghan, ``Elimination of electrocardiogram
  contamination from electromyogram signals: An evaluation of currently used
  removal techniques,'' \emph{Journal of electromyography and kinesiology},
  vol.~16, no.~2, pp. 175--187, 2006.

\bibitem{aminian1988filtering}
K.~Aminian, C.~Ruffieux, and P.~Robert, ``Filtering by adaptive sampling
  (fas),'' \emph{Medical and Biological Engineering and Computing}, vol.~26,
  no.~6, pp. 658--662, 1988.

\bibitem{abbaspour2014removing}
S.~Abbaspour and A.~Fallah, ``Removing ecg artifact from the surface emg signal
  using adaptive subtraction technique,'' \emph{Journal of biomedical physics
  \& engineering}, vol.~4, no.~1, p.~33, 2014.

\bibitem{nougarou2018efficient}
F.~Nougarou, D.~Massicotte, and M.~Descarreaux, ``Efficient procedure to remove
  ecg from semg with limited deteriorations: Extraction, quasi-periodic
  detection and cancellation,'' \emph{Biomedical Signal Processing and
  Control}, vol.~39, pp. 1--10, 2018.

\bibitem{abbaspour2015ecg}
S.~Abbaspour, M.~Lind{\'e}n, and H.~Gholamhosseini, ``Ecg artifact removal from
  surface emg signal using an automated method based on wavelet-ica.'' in
  \emph{pHealth}, 2015, pp. 91--97.

\bibitem{chen2016fastica}
M.~Chen, X.~Zhang, X.~Chen, M.~Zhu, G.~Li, and P.~Zhou, ``Fastica peel-off for
  ecg interference removal from surface emg,'' \emph{Biomedical engineering
  online}, vol.~15, no.~1, p.~65, 2016.

\bibitem{li2013ecg}
Y.~Li, X.~Chen, X.~Zhang, and P.~Zhou, ``Ecg artifact removal from emg
  recordings using independent component analysis and adapted filter,'' in
  \emph{Neural Engineering (NER), 2013 6th International IEEE/EMBS Conference
  on}.\hskip 1em plus 0.5em minus 0.4em\relax IEEE, 2013, pp. 347--350.

\bibitem{akselrod1981power}
S.~Akselrod, D.~Gordon, F.~A. Ubel, D.~C. Shannon, A.~Berger, and R.~J. Cohen,
  ``Power spectrum analysis of heart rate fluctuation: a quantitative probe of
  beat-to-beat cardiovascular control,'' \emph{science}, vol. 213, no. 4504,
  pp. 220--222, 1981.

\bibitem{barlaam2011time}
F.~Barlaam, M.~Descoins, O.~Bertrand, T.~Hasbroucq, F.~Vidal, C.~Assaiante, and
  C.~Schmitz, ``Time--frequency and erp analyses of eeg to characterize
  anticipatory postural adjustments in a bimanual load-lifting task,''
  \emph{Frontiers in human neuroscience}, vol.~5, p. 163, 2011.

\bibitem{krigolson2017event}
O.~E. Krigolson, ``Event-related brain potentials and the study of reward
  processing: Methodological considerations,'' \emph{International Journal of
  Psychophysiology}, 2017.

\bibitem{cavanagh1979electromechanical}
P.~R. Cavanagh and P.~V. Komi, ``Electromechanical delay in human skeletal
  muscle under concentric and eccentric contractions,'' \emph{European journal
  of applied physiology and occupational physiology}, vol.~42, no.~3, pp.
  159--163, 1979.

\bibitem{bonato1998statistical}
P.~Bonato, T.~D'Alessio, and M.~Knaflitz, ``A statistical method for the
  measurement of muscle activation intervals from surface myoelectric signal
  during gait,'' \emph{IEEE Transactions on biomedical engineering}, vol.~45,
  no.~3, pp. 287--299, 1998.

\bibitem{drapala2012two}
J.~Drapa{\l}a, K.~Brzostowski, A.~Szpala, and A.~Rutkowska-Kucharska, ``Two
  stage emg onset detection method,'' \emph{Archives of Control Sciences},
  vol.~22, no.~4, pp. 427--440, 2012.

\end{thebibliography}

\end{document}